\documentclass[12pt]{article}
\usepackage{epsf}
\newcommand{\be}{\begin{equation}}
\newcommand{\ee}{\end{equation}}
\newcommand{\bea}{\begin{eqnarray}}
\newcommand{\eea}{\end{eqnarray}}
\newcommand{\ba}{\begin{array}}
\newcommand{\ea}{\end{array}}
\newcommand{\pl}{Phys.\ Lett.}
\newcommand{\np}{Nucl.\ Phys.}

\title{ {\bf
$B_s\rightarrow \gamma \gamma$ decay in the two Higgs doublet model with flavor 
changing neutral currents}}
\author{\vspace{1cm}\\
         {\bf T. M. Aliev} , \\
        Physics Department, Girne American University \\
        Girne , Mersin-10, Turkey \\
        \vspace{5mm}\\
        {\bf E. O. Iltan}
        \thanks{E-mail address:
        eiltan@heraklit.physics.metu.edu.tr}
 \\
        Physics Department, Middle East Technical University \\
        Ankara, Turkey\\}

\date{}

\begin{document}
\setlength{\baselineskip}{24pt}
\maketitle
\setlength{\baselineskip}{7mm}
\begin{abstract}
We calculate the leading logarithmic QCD corrections to the decay
$B_{s}\rightarrow\gamma\gamma$ in the two Higgs doublet model with tree
level flavor changing currents (model III) including $O_{7}$ type long 
distance effects. Further, we analyse the dependencies of the branching 
ratio $Br(B_{s}\rightarrow\gamma\gamma)$ and the ratio of CP-even and 
CP-odd amplitude squares, $R=|A^{+}|^2/|A^{-}|^2$, on the charged Higgs 
mass $m_{H^{\pm}}$ and the selected model III parameters $\xi^{U,D}$ 
including the leading logarithmic QCD corrections. It is found that to 
look for charged Higgs effects, the measurement of the branching ratio 
$Br(B_{s}\rightarrow\gamma\gamma)$ is promising.
\end{abstract} 
\thispagestyle{empty}
\newpage
\setcounter{page}{1}

\section{Introduction}
Rare B meson decays constitute one of the most important classes of decays
since they are induced by flavor changing neutral currents (FCNC) at loop 
level in the Standard Model (SM). Therefore they allow us to test the flavor 
structure of the SM and provide a comprehensive information about the
fundamental parameters, such as Cabbibo-Kobayashi-Maskawa (CKM) 
matrix elements, leptonic decay constants, CP ratio, etc.  
These decays  are also sensitive to the new physics beyond the SM, such as
two Higgs Doublet model (2HDM), Minimal Supersymmetric extension of the SM
(MSSM) \cite{Hewett}, etc.

Among the rare B meson decays, the exclusive $B_s\rightarrow \gamma \gamma$ 
decay has received considerable interest in view of the planned experiments
at the upcoming KEK and SLAC-B factories and existing hadronic accelerators,
which may measure the branching ratios ($Br$) as low as $10^{-8}$.
Since the $B_s\rightarrow \gamma\gamma$ decay has two photon system, it is
possible to study the CP violating effects \cite{kalinowski} and 
it can be easily detected in the experiments by putting a cut for the energy 
of final photons \cite{Ricciardi}. Further, this decay can give information 
about physics beyond the SM.

The $B_s\rightarrow \gamma\gamma$ decay  has been studied in the framework 
of the SM (2HDM) \cite{kalinowski},\cite{yao}-\cite{simma} (\cite{aliev}) 
without QCD corrections. It is well-known, that the QCD corrections to the  
$b\rightarrow s\gamma$ decay are considerably large (see \cite{Grinstein} 
- \cite{burasmisiak} and references therein).
Therefore, one can naturally expect that the situation is the same 
for the inclusive $b\rightarrow s \gamma\gamma$ decay. 
Recently, the analysis with the addition of the leading logarithmic (LLog) 
QCD corrections in the SM \cite{Gud}-\cite{Soni}, 2HDM 
\cite{gudalil} and MSSM \cite{bertolini} has been done and the strong 
sensitivity to QCD corrections was obtained.

The calculations shows that the $Br$ of the decay 
$B_s\rightarrow \gamma\gamma$ is enhanced with the addition of the QCD 
corrections in the SM. The extension of the Higgs sector (model II in 2HDM) 
brings extra enhancement to the $Br$. However, the present theoretical results are at the order of 
($\sim 10^{-6}$) \cite{gudalil} are far from the experimental upper limit \cite{Acciarri}
\begin{eqnarray}
Br (B_s\rightarrow \gamma\gamma) \le 1.48\, . 10^{-4} \,\, .
\label{br1}
\end{eqnarray}

In the present work, we study this decay in the framework of the 
two Higgs doublet model with three level flavor changing neutral currents
(model III) including perturbative QCD corrections in the LLog approximation,  
by imposing a method based on heavy quark effective theory for the bound 
state of the $B_s$ meson \cite{Gud}.
Further, we improve our calculations with the inclusion of long-distance 
effects through the transition $B_s \to \phi \gamma \to \gamma \gamma$, 
which we call $O_7$-type, see \cite{Gud} for details. 
In the analysis, we use the constraints coming from 
$\Delta F=2 \, (F=K,D,B)$ mixing, the $\rho$ parameter, the ratio 
$R_b^{exp}=\Gamma(Z\rightarrow b\bar{b})/\Gamma(Z\rightarrow hadrons)
\, (=0.2178\pm0.0011)$ and 
the CLEO measurement of the decay $B\rightarrow X_s\gamma$ \cite{cleo},
for the selected parameters of model III \cite{alil},  
and we get an extreme enhancement for the $Br ( B_s\rightarrow
\gamma\gamma)$. We also predict an upper limit for the model III
parameter $\bar{\xi}_{bb}^{D}$ using the present experimental restriction 
for the $Br$ of the decay $B_s\rightarrow\gamma\gamma$, eq.~(\ref{br1}).    

The paper is organized as follows:
In section 2, we give the LLog QCD corrected amplitude for the exclusive 
decay $B_s\rightarrow \gamma\gamma$.
We further calculate the CP-odd $A^-$ and CP-even $A^+$ amplitudes in 
an approach based on heavy quark effective theory, taking the LLog QCD 
corrections into account. Section 3 is devoted to the analysis of the
dependencies of the $Br$ and the ratio $|A^{+}|^2/|A^{-}|^2$
on the parameters $\mu$ (scale parameter), the Yukawa couplings 
$\bar{\xi}_{bb}^{D}$, $\bar{\xi}_{tt}^{U}$ and our conclusions.
In the appendix , we present the operator basis and the Wilson coefficients 
responsible for the inclusive $b\rightarrow s\gamma\gamma$ decay in the
model III. We also discuss the contributions of the neutral Higgs bosons to 
the coefficient $C_7$ for the decay $b\rightarrow s\gamma$ and the 
restrictions to the free parameters $\bar{\xi}_{ij}^{D}$, where one of 
the indices $i$ or $j$ belong to the first or second generation. 
\section{\bf Leading logarithmic improved short-distance contributions in 
the model III for the decay $B_s\rightarrow \gamma \gamma $ }
In this section we present the effective Hamiltonian 
to the exclusive $B_s\rightarrow  \gamma \gamma $ decay amplitude in the 
2HDM with tree level neutral currents (model III). 
We start with the brief explanation of the model under consideration. 
The Yukawa interaction in the general case is
\begin{eqnarray}
{\cal{L}}_{Y}=\eta^{U}_{ij} \bar{Q}_{i L} \tilde{\phi_{1}} U_{j R}+
\eta^{D}_{ij} \bar{Q}_{i L} \phi_{1} D_{j R}+
\xi^{U}_{ij} \bar{Q}_{i L} \tilde{\phi_{2}} U_{j R}+
\xi^{D}_{ij} \bar{Q}_{i L} \phi_{2} D_{j R} + h.c. \,\,\, ,
\label{lagrangian}
\end{eqnarray}
where $L$ and $R$ denote chiral projections $L(R)=1/2(1\mp \gamma_5)$,
$\phi_{i}$ for $i=1,2$, are the two scalar doublets, $\eta^{U,D}_{ij}$
and $\xi^{U,D}_{ij}$ are the matrices of the Yukawa couplings.
The choice of $\phi_1$ and $\phi_2$
\begin{eqnarray}
\phi_{1}=\frac{1}{\sqrt{2}}\left[\left(\begin{array}{c c} 
0\\v+H^{0}\end{array}\right)\; + \left(\begin{array}{c c} 
\sqrt{2} \chi^{+}\\ i \chi^{0}\end{array}\right) \right]\, ; 
\phi_{2}=\frac{1}{\sqrt{2}}\left(\begin{array}{c c} 
\sqrt{2} H^{+}\\ H_1+i H_2 \end{array}\right) \,\, ,
\label{choice}
\end{eqnarray}
with the vacuum expectation values,  
\begin{eqnarray}
<\phi_{1}>=\frac{1}{\sqrt{2}}\left(\begin{array}{c c} 
0\\v\end{array}\right) \,  \, ; 
<\phi_{2}>=0 \,\, ,
\label{choice2}
\end{eqnarray}
permits us to write the Flavor Changing (FC) part of the interaction as 
\begin{eqnarray}
{\cal{L}}_{Y,FC}=
\xi^{U}_{ij} \bar{Q}_{i L} \tilde{\phi_{2}} U_{j R}+
\xi^{D}_{ij} \bar{Q}_{i L} \phi_{2} D_{j R} + h.c. \,\, ,
\label{lagrangianFC}
\end{eqnarray}
where the couplings  $\xi^{U,D}$ for the FC charged interactions are
\begin{eqnarray}
\xi^{U}_{ch}&=& \xi_{neutral} \,\, V_{CKM} \nonumber \,\, ,\\
\xi^{D}_{ch}&=& V_{CKM} \,\, \xi_{neutral} \,\, ,
\label{ksi1} 
\end{eqnarray}
and  $\xi^{U,D}_{neutral}$ 
\footnote{In all next discussion we denote $\xi^{U,D}_{neutral}$ 
as $\xi^{U,D}_{N}$.} 
is defined by the expression
\begin{eqnarray}
\xi^{U,D}_{N}=(V_L^{U,D})^{-1} \xi^{U,D} V_R^{U,D}\,\, .
\label{ksineut}
\end{eqnarray}
Here, the charged couplings appear as a linear combinations of neutral 
couplings multiplied by $V_{CKM}$ matrix elements. 

Now, we would like to discuss the LLog QCD corrections to 
$b\rightarrow s\gamma\gamma$ decay amplitude in the model III. 
As it is well known, the effective Hamiltonian method, obtained by 
integrating out the heavy degrees of freedom, is a powerfull one.
In the present case, $t$ quark, $W^{\pm}, H^{\pm}, H_{1}$, and $H_{2}$ 
bosons are the heavy degrees of freedom. Here $H^{\pm}$ denote charged, 
$H_{1}$ and $H_{2}$ denote neutral Higgs bosons. The LLog QCD corrections 
are done through matching the full theory with the effective low energy 
theory at the high scale $\mu=m_{W}$ and evaluating the Wilson coefficients 
from $m_{W}$ down to the lower scale $\mu\sim O(m_{b})$.
Note that we choose the higher scale as $\mu=m_{W}$ since the evaluation 
from the scale $\mu=m_{H^{\pm}}$ to $\mu=m_{W}$ gives negligible
contribution to the Wilson coefficients ($\sim 5\%$) since the charged 
Higgs boson is heavy due to the current theoretical restrictions based on
the experimental measurement of $B\rightarrow X_s\gamma$ due to the CLEO
collaboration \cite{cleo}, for example, 
$m_H^{\pm} \geq 340\, GeV$ \cite{ciuchini},
$m_H^{\pm} \geq 480\, GeV$ \cite{gudalil}.
      
The effective Hamiltonian relevant for our process is
\begin{eqnarray}
{\cal{H}}_{eff}=-4 \frac{G_{F}}{\sqrt{2}} V_{tb} V^{*}_{ts} 
\sum_{i}C_{i}(\mu) O_{i}(\mu) \, \, ,
\label{hamilton}
\end{eqnarray}
where the $O_{i}$ are operators given in eqs.~(\ref{op1}),~(\ref{op2}) 
and the $C_{i}$ are Wilson coefficients
renormalized at the scale $\mu$. The coefficients $C_{i}$ are calculated 
perturbatively. The explicit forms of the full-set operators and the
corresponding Wilson coefficients are presented in Appendix.

To obtain the decay amplitude for the $B_s\rightarrow \gamma\gamma$
decay, we need to sandwich the effective Hamiltonian between
the $B_s$ and two photon states, i.e. $<B_s| {\cal{H}}_{eff}|\gamma\gamma>$.
This matrix element can be written in terms of two Lorentz structures 
\cite{simma} - \cite{aliev}, \cite{Gud}-\cite{Soni}:  
\begin{equation}
{\cal A}(B_{s}\rightarrow \gamma \gamma)=
A^{+} {\cal F}_{\mu\nu} {\cal F}^{\mu\nu} +
i A^{-} {\cal F}_{\mu\nu} \tilde{{\cal F}}^{\mu\nu}
\, \, ,
\label{amp}
\end{equation}
where 
$\tilde{{\cal F}}_{\mu\nu}=\frac{1}{2}\epsilon_{\mu\nu\alpha\beta} 
{\cal F}^{\alpha\beta}$.
Here, the two different operator sets (eqs. (\ref{op1}) and (\ref{op2})) give
contributions to the CP-even $A^{+}$ and CP-odd $A^{-}$ parts.
We denote the CP-even (CP-odd) amplitudes due to the first set as 
$A_1^{+}$ ($A_1^{-}$) and the second set as $A_2^{+}$ ($A_2^{-}$).
In a HQET inspired approach these amplitudes are:
\begin{eqnarray}
A_1^{+}&=&\frac{\alpha_{em} G_F}{\sqrt{2} \pi} \frac{f_{B_s}}{m_{B_{s}}^{2}} 
\lambda_t 
\left( \frac{1}{3}
\frac{m^4_{B_s} (m_b^{eff}-m_s^{eff})}{\bar{\Lambda}_s 
(m_{B_s}-\bar{\Lambda}_s) (m_b^{eff}+m_s^{eff})} 
C_7^{eff}(\mu)
\nonumber
\right.\\
&-&
\left.
\frac{4}{9} \frac{m_{B_{s}^2}}{m_b^{eff}+m_s^{eff}}\, [ \,
(-m_b J(m_b)+ m_s J(m_s) ) D(\mu) - m_c J(m_c)  E(\mu)\, ]\,  
\right)  ,\nonumber\\
A_1^{-}&=&- \frac{\alpha_{em} G_F}{\sqrt{2} \pi} f_{B_s} \lambda_t 
\left( \frac{1}{3}
\frac{1}{m_{B_s} \bar{\Lambda}_s (m_{B_s}-\bar{\Lambda}_s)} g_{-}
C_7^{eff}(\mu) -\sum_q Q_q^2 I(m_q) C_q(\mu) \nonumber 
\right.\\
&+&
\left.
\frac{1}{9 (m_b^{eff}+m_s^{eff})} 
[\, (m_b \triangle(m_b)+m_s \triangle(m_s)) D(\mu)+ m_c \triangle(m_c)
E(\mu) \,] \right) \, \, ,
\label{Amplitudes1}
\end{eqnarray}
and 
\begin{eqnarray}
A_2^{+}&=&-\frac{\alpha_{em} G_F}{\sqrt{2} \pi} \frac{f_{B_s}}{m_{B_{s}}^{2}} 
\lambda_t 
\left( \frac{1}{3}
\frac{m^4_{B_s} (m_b^{eff}-m_s^{eff})}{\bar{\Lambda}_s 
(m_{B_s}-\bar{\Lambda}_s) (m_b^{eff}+m_s^{eff})} C_7^{\prime eff}(\mu) \right )
\nonumber \,\, , \\
A_2^{-}&=&- \frac{\alpha_{em} G_F}{\sqrt{2} \pi} f_{B_s} \lambda_t 
\left( \frac{1}{3}
\frac{1}{m_{B_s} \bar{\Lambda}_s (m_{B_s}-\bar{\Lambda}_s)} g_{-}
C_7^{\prime eff}(\mu) \right ) \,\, ,
\label{Amplitudes2}
\end{eqnarray}
where $Q_q=\frac{2}{3}$ for $q=u,c$ and $Q_q=-\frac{1}{3}$ for $q=d,s,b$.
Here, we have used the unitarity of the CKM-matrix 
$\sum_{i=u,c,t} V_{is}^{*} V_{ib}=0 $ 
and have neglected the contribution due to 
$V_{us}^{*} V_{ub} \ll V_{ts}^{*} V_{tb}\equiv \lambda_t$. 
The function $g_{-}$ is defined as \cite{Gud}:
\begin{eqnarray}
g_{-}=m_{B_s}(m_b^{eff}+m_s^{eff})^2+
\bar{\Lambda}_s (m^2_{B_s}-(m_b^{eff}+m_s^{eff})^2) 
\,\, .
\label{gmin}
\end{eqnarray}
The parameter $\bar{\Lambda}_{s}$ enters eq.~(\ref{Amplitudes1}) and 
~(\ref{Amplitudes2}) through the bound state kinematics \cite{Gud}.
Now, we would like explain the approach we follow to be clear how to get the 
amplitudes ( eqs. ~(\ref{Amplitudes1}) and ~(\ref{Amplitudes2})).
The momentum of the $\bar{b}-$quark inside the $B_s$ meson can be written as 
$p=m_b v+k$, where $k$ is the residual momentum and $v$ is the 4-velocity
defined by $v=p_B/m_{B_s}$. Here $p_B$ is the 4-momentum of the meson
$B_s$. In the matrix elements we need to evaluate $p.k_i$ and $p'.k_i$, 
$i=1,2$, where $k_i$, $p'$ are the momenta of the outgoing photon and $s$
-quark respectively. We average the residual momentum of the $\bar{b}$ 
quark as \cite{MW}
\begin{eqnarray}
<k_{\mu}>=-\frac{1}{2 m_b}(\lambda_1+3 \lambda_2) v_{\mu} 
\nonumber \,\, , \\
<k_{\mu} k_{\nu}>=\frac{\lambda_1}{3}(g_{\mu\nu}-v_{\mu} v_{\nu})\,\, ,
\end{eqnarray}
where $\lambda_1$, $\lambda_2$ are matrix elements from heavy quark 
expansion. $\lambda_2\, (=0.12\, GeV)$ can be determined from the 
$B^*_{(s)}-B_{(s)}$ mass splitting. 

Finally we have
\begin{eqnarray}
p.k_i&=&\frac{m_{B_s}}{2}(m_{B_s}-\bar{\Lambda}_{s})\nonumber \,\, ,\\
p'.k_i&=&-\frac{m_{B_s}}{2}\bar{\Lambda}_{s}\nonumber \,\, ,\\
(m_{b}^{eff})^{2}&=& p^2=m_{b}^{2}-3 \lambda_{2}\,\, , \nonumber \\
(m_{s}^{eff})^{2}&=& p'^2=(m_{s}^{eff})^{2}-m_{B_{s}}^{2}+
2 m_{B_{s}}\bar{\Lambda}_{s}\,\, . 
\label{HQET}
\end{eqnarray}
Here, we use the definitions $p_B=p-p'$, $p_B.k_i=\frac{m_{B_s}^2}{2}$, 
$v.k_i=\frac{m_{B_s}}{2}$ and the HQET relation
\cite{MW}
\begin{eqnarray}
m_{B_s}=m_b+\bar{\Lambda}_{s}-\frac{1}{2 m_b}(\lambda_1+3 \lambda_2)\,\, ,
\label{mBs}
\end{eqnarray}
In eq. (\ref{HQET}), $m_b^{eff}$ and $m_s^{eff}$ are the effective masses
of the quarks in the $B_{s}$ meson bound state \cite{Gud} and 
the parameter $\bar{\Lambda}_{s}$ can be defined as
$\bar{\Lambda}_{s}=\bar{\Lambda}+\Delta m$. $\bar{\Lambda}$
together with $\lambda_1$ can be extracted from data on semileptonic 
$B^{\pm}\, , B_0$ decays \cite{gremm} and the measured mass difference 
$\Delta m=m_{B_s}-m_B=90\, MeV$ \cite{barnett}. Using eq. (\ref{mBs}),
we correlate the parameters $\bar{\Lambda}_{s}$ and $m_b$ and 
$\bar{\Lambda}$ and $\lambda_1$ (Table (~\ref{input})).

After this brief explanation of the HQET inspired approach we follow, 
we continue with the LLog QCD-corrected Wilson 
coefficients $C_{1 \dots 10}(\mu)$ \cite{Gud} - \cite{Soni}
which enter the amplitudes in the combinations
\begin{eqnarray}
C_u(\mu)&=&C_d(\mu)=(C_3(\mu)-C_5(\mu)) N_c +C_4(\mu)-C_6(\mu) \, \, , 
\nonumber \\
C_c(\mu)&=&
(C_1(\mu)+C_3(\mu)-C_5(\mu)) N_c +C_2(\mu)+C_4(\mu)-C_6(\mu) \, \, ,
\nonumber \\
C_s(\mu)&=&C_b(\mu)=
(C_3(\mu)+C_4(\mu))(N_c+1)-N_c C_5(\mu)-C_6(\mu) \, \, , \nonumber \\
D(\mu)&=&C_5(\mu)+C_6(\mu) N_c \, \, , \nonumber \\
E(\mu)&=&C_{10}(\mu)+C_9(\mu) N_c
\end{eqnarray} 
where $N_c$ is the number of colours ($N_c=3$ for QCD).

The effective coefficients $C_7^{eff}(\mu)$ and $C_7^{\prime eff}(\mu)$ 
are defined in the NDR scheme, which we use here, as \cite{alil}
\begin{eqnarray}
C_{7}^{eff}(\mu)&=&C_{7}^{2HDM}(\mu)+ Q_d \, 
(C_{5}^{2HDM}(\mu) + N_c \, C_{6}^{2HDM}(\mu))\nonumber \, \, , \\
&+& Q_u\, (\frac{m_c}{m_b}\, C_{10}^{2HDM}(\mu) + N_c \, 
\frac{m_c}{m_b}\,C_{9}^{2HDM}(\mu))\nonumber \, \, , \\
C^{\prime eff}_7(\mu)&=& C^{\prime 2HDM}_7(\mu)+Q_{d}\, 
(C^{\prime 2HDM}_5(\mu) + N_c \, C^{\prime 2HDM}_6(\mu))\nonumber \\
&+& Q_u (\frac{m_c}{m_b}\, C_{10}^{\prime 2HDM}(\mu) + N_c \, 
\frac{m_c}{m_b}\,C_{9}^{\prime 2HDM}(\mu))\, \, .
\label{C7eff}
\end{eqnarray}
The functions $I(m_q), \, J(m_q)$ and $\triangle(m_q)$ come from the 
irreducible diagrams with an internal 
$q$ type quark propagating and are defined as
\begin{eqnarray}
I(m_q)&=&1+\frac{m_q^2}{m_{B_s}^2} \triangle (m_q) \, \, , \nonumber \\
J(m_q)&=&1-\frac{m_{B_s}^2-4 m_q^2}{4 m_{B_s}^2} \triangle(m_q)  \, \, ,
\nonumber \\
\triangle(m_q)&=&\left(
\ln(\frac{m_{B_s}+\sqrt{m_{B_s}^2-4 m_q^2}}
         {m_{B_s}-\sqrt{m_{B_s}^2-4 m_q^2}})-i \pi \right)^2 
\, \,{\mbox{for}}\, \, \frac{m^2_{B_s}}{4 m_q^2} \geq 1 , \nonumber \\
\triangle(m_q)&=&-\left(
2 \arctan(\frac{\sqrt{4 m_q^2-m_{B_s}^2}}
         {m_{B_s}})-\pi \right)^2 
\, \,{\mbox{for}}\, \, \frac{m^2_{B_s}}{4 m_q^2} <\ 1 .
\end{eqnarray}
Finally the CP-even $A^-$ and CP-odd $A^+$ amplitudes can be written as
\begin{eqnarray}
A^{+}&=& A_{1}^{+}+A_{2}^{+} \nonumber \,\, , \\
A^{-}&=& A_{1}^{-}+A_{2}^{-} \,\, ,
\label{finamp}
\end{eqnarray}
where the amplitudes $A_{1}^{\pm}$ and $A_{2}^{\pm}$ are given in eqs.
~(\ref{Amplitudes1}) and ~(\ref{Amplitudes2}) respectively.

In our numerical analysis we used the input values given in 
Table~(\ref{input}).
\begin{table}[h]
        \begin{center}
        \begin{tabular}{|l|l|}
        \hline
        \multicolumn{1}{|c|}{Parameter} & 
                \multicolumn{1}{|c|}{Value}     \\
        \hline \hline
        $m_c$                   & $1.4$ (GeV) \\
        $m_b$                   & $4.8$ (GeV) \\
        $\alpha_{em}^{-1}$      & 129           \\
        $\lambda_t$            & 0.04 \\
        $\Gamma_{tot}(B_s)$             & $4.09 \cdot 10^{-13}$ (GeV)   \\
        $f_{B_s}$             & $0.2$ (GeV)  \\   
        $m_{B_s}$             & $5.369$ (GeV) \\
        $m_{t}$             & $175$ (GeV) \\
        $m_{W}$             & $80.26$ (GeV) \\
        $m_{Z}$             & $91.19$ (GeV) \\
        $\Lambda^{(5)}_{QCD}$             & $0.214$ (GeV) \\
        $\alpha_{s}(m_Z)$             & $0.117$  \\
        $\lambda_2$             & $0.12$ $(\mbox{GeV}^2)$ \\
        $\lambda_1$             & $-0.29$ $(\mbox{GeV}^2)$ \\ 
        $\bar{\Lambda}_s$             & $590$ $(\mbox{MeV})$ \\
        $\bar{\Lambda}$             & $500$ $(\mbox{MeV})$ \\
        \hline
        \end{tabular}
        \end{center}
\caption{Values of the input parameters used in the numerical
          calculations unless otherwise specified.}
\label{input}
\end{table}
\newpage
\section{Discussion}
Before we present our numerical results, we would like to discuss briefly 
the free parameters of the model under consideration, i.e. model III.
This model induces many free parameters, such as $\xi_{ij}^{U,D}$ where 
i,j are flavor indices. To obtain qualitative results, we should restrict
them using the experimental measurements. 
The explicit expressions of $C_7^{h_0}$ and $C_7^{A_0}$ (eq. (\ref{c7A0h0})) 
show that the the neutral Higgs bosons can give a large contribution to the 
coefficient $C_7 $ which can be in contradiction with the CLEO data 
\cite{cleo}, 
\begin{eqnarray}
Br (B\rightarrow X_s\gamma)= (2.32\pm0.07\pm0.35)\, 10^{-4} \,\, .
\label{br2}
\end{eqnarray}
Such potentially dangerous terms are removed with the assumption that the 
couplings $\bar{\xi}^{D}_{N,is}$($i=d,s,b)$ and $\bar{\xi}^{D}_{N,db}$ are 
negligible to be able to reach the conditions 
$\bar{\xi}^{D}_{N,bb} \,\bar{\xi}^{D}_{N,is} <<1$ and 
$\bar{\xi}^{D}_{N,db} \,\bar{\xi}^{D}_{N,ds} <<1$.
Further, we use the constraints \cite{alil} coming from the 
ratio $R_b^{exp}=\Gamma(Z\rightarrow b\bar{b})/\Gamma(Z\rightarrow hadrons)$, 
namely $\xi^{D}_{N bb} > \frac{60 m_b}{v}$, the restrictions due 
to the $\Delta F=2$ mixing, the $\rho$ parameter \cite{atwood}, 
and the measurement by CLEO collaboration. 
The analysis of the mentioned processes and the discussion given above, 
leads to choose  $\bar{\xi}_{N tc} << \bar{\xi}^{U}_{N tt},
\bar{\xi}^{D}_{N bb}$ and $\bar{\xi}^{D}_{N ib} \sim 0\, , 
\bar{\xi}^{D}_{N ij}\sim 0$, where the indices $i,j$ denote d and s quarks . 

\begin{figure}[htb]
\vskip -1.5truein
\centering
\epsfxsize=3.8in
\leavevmode\epsffile{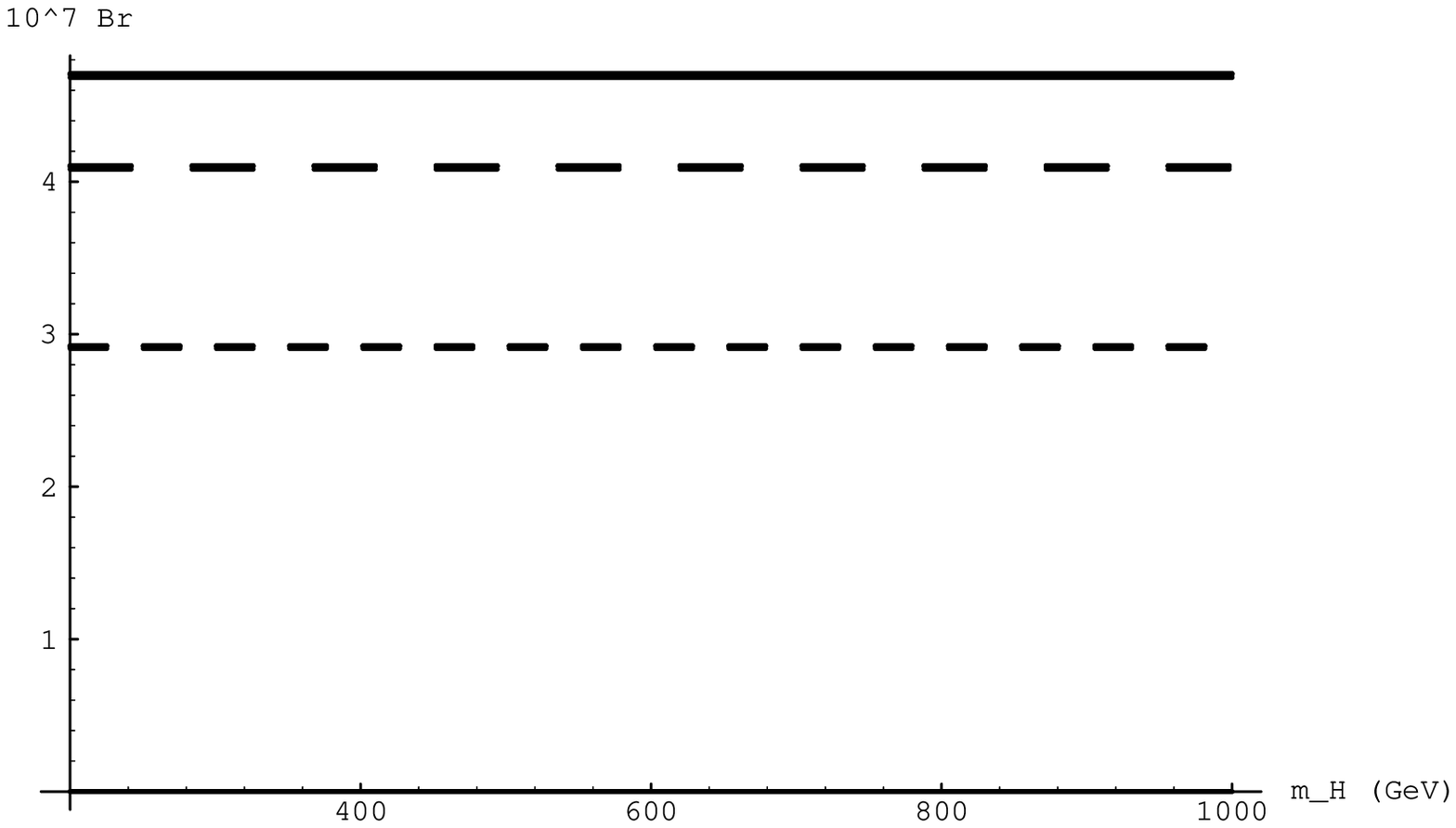}
\vskip -1.5truein
\caption[]{$Br$ as a function of the mass $m_{H^{\pm}}$ 
for fixed $\bar{\xi}_{N,bb}^{D}=60\, m_b$ at the region $|r_{tb}|<<1$.
Here solid lines correspond to the scale $\mu=2.5\,\, GeV$, 
dashed lines to $\mu=5\,\, GeV$ and small dashed lines to $\mu=m_W\,\, GeV$. 
The lines corresponding to the SM coincides with the lines we present here.}
\label{brbb603mha}
\end{figure}

\begin{figure}[htb]
\vskip -1.5truein
\centering
\epsfxsize=3.8in
\leavevmode\epsffile{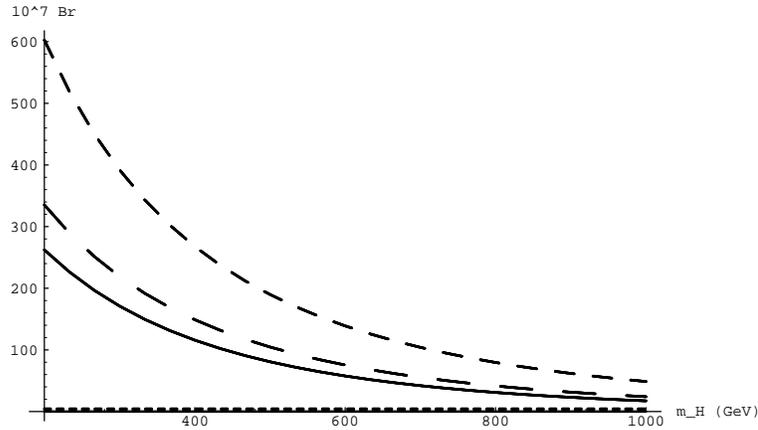}
\vskip -1.5truein
\caption[]{The same as Fig 1, but at the region $r_{tb} >> 1$. Dotted dashed
lines correspond to the SM. Note that dotted dashed line almost coincides
with the $m_H$ axis}
\label{brbb603mhb}
\end{figure}

\begin{figure}[htb]
\vskip -1.5truein
\centering
\epsfxsize=3.8in
\leavevmode\epsffile{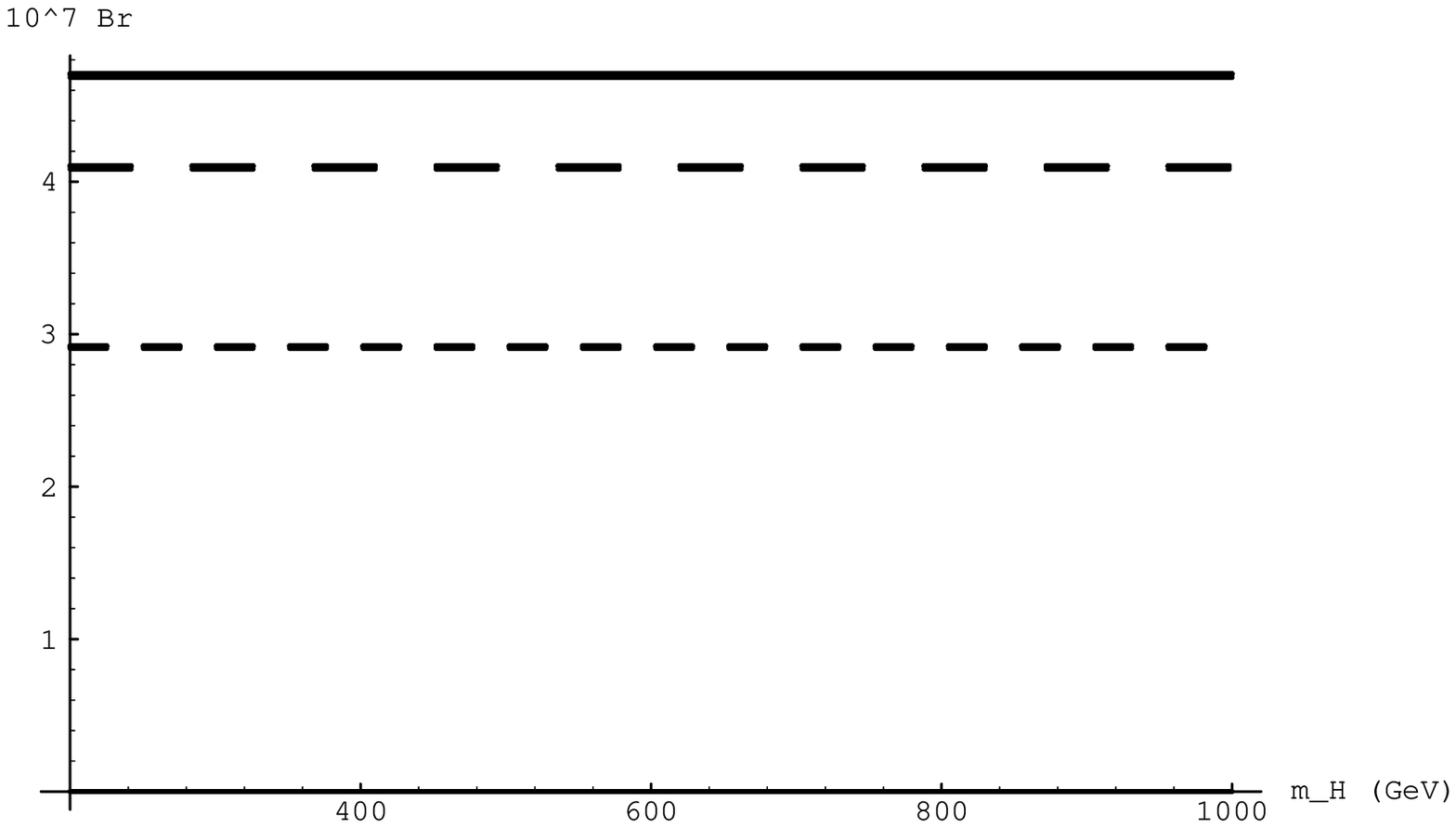}
\vskip -1.5truein
\caption[]{The same as Fig 1, but for fixed $\bar{\xi}_{N,bb}^{D}=90\, m_b$ 
value.}
\label{brbb903mha}
\end{figure}

\begin{figure}[htb]
\vskip -1.5truein
\centering
\epsfxsize=3.8in
\leavevmode\epsffile{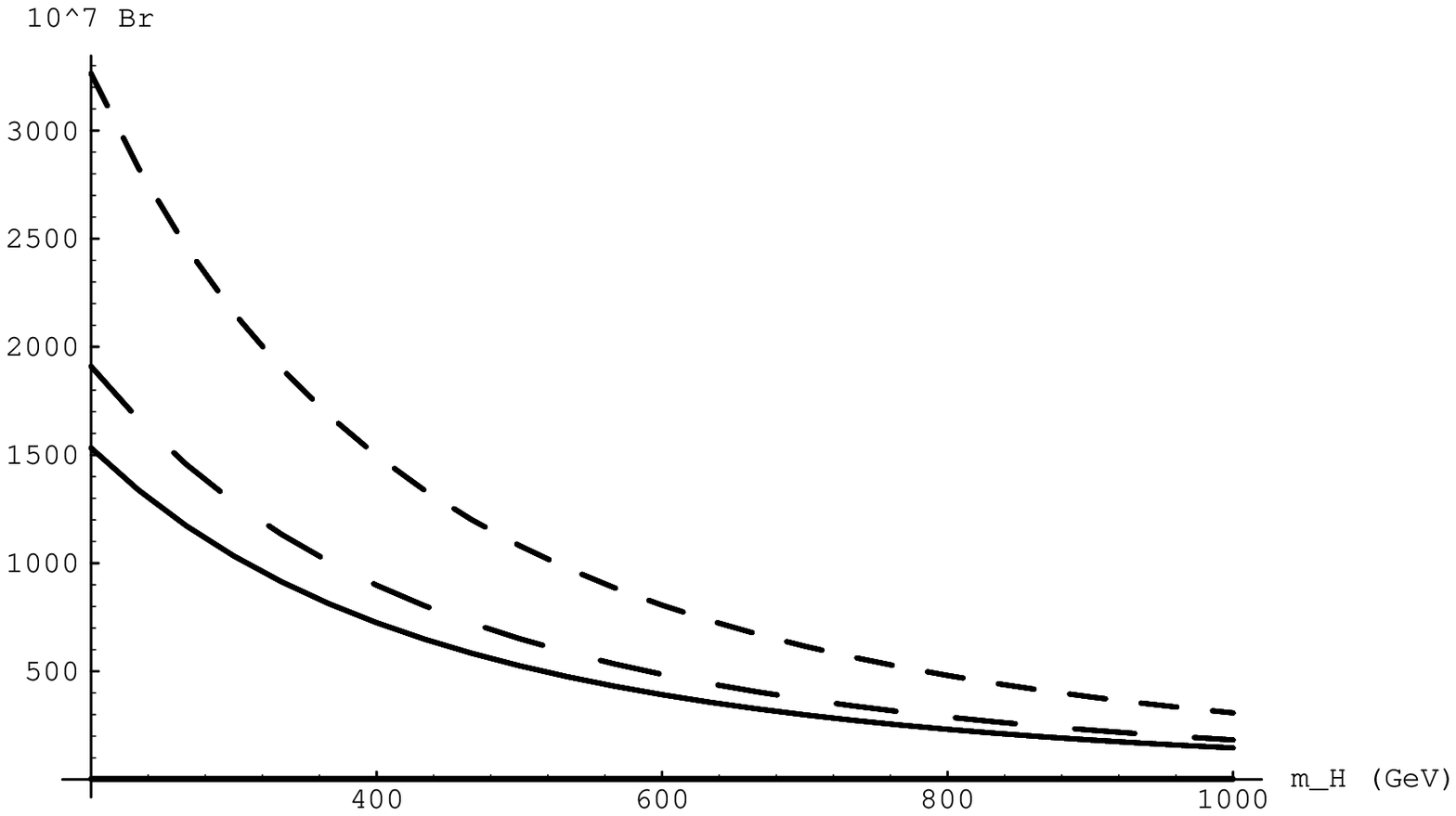}
\vskip -1.5truein
\caption[]{The same as Fig 2, but for fixed $\bar{\xi}_{N,bb}^{D}=90\, m_b$ 
value.}
\label{brbb903mhb}
\end{figure}

\begin{figure}[htb]
\vskip -1.5truein
\centering
\epsfxsize=3.8in
\leavevmode\epsffile{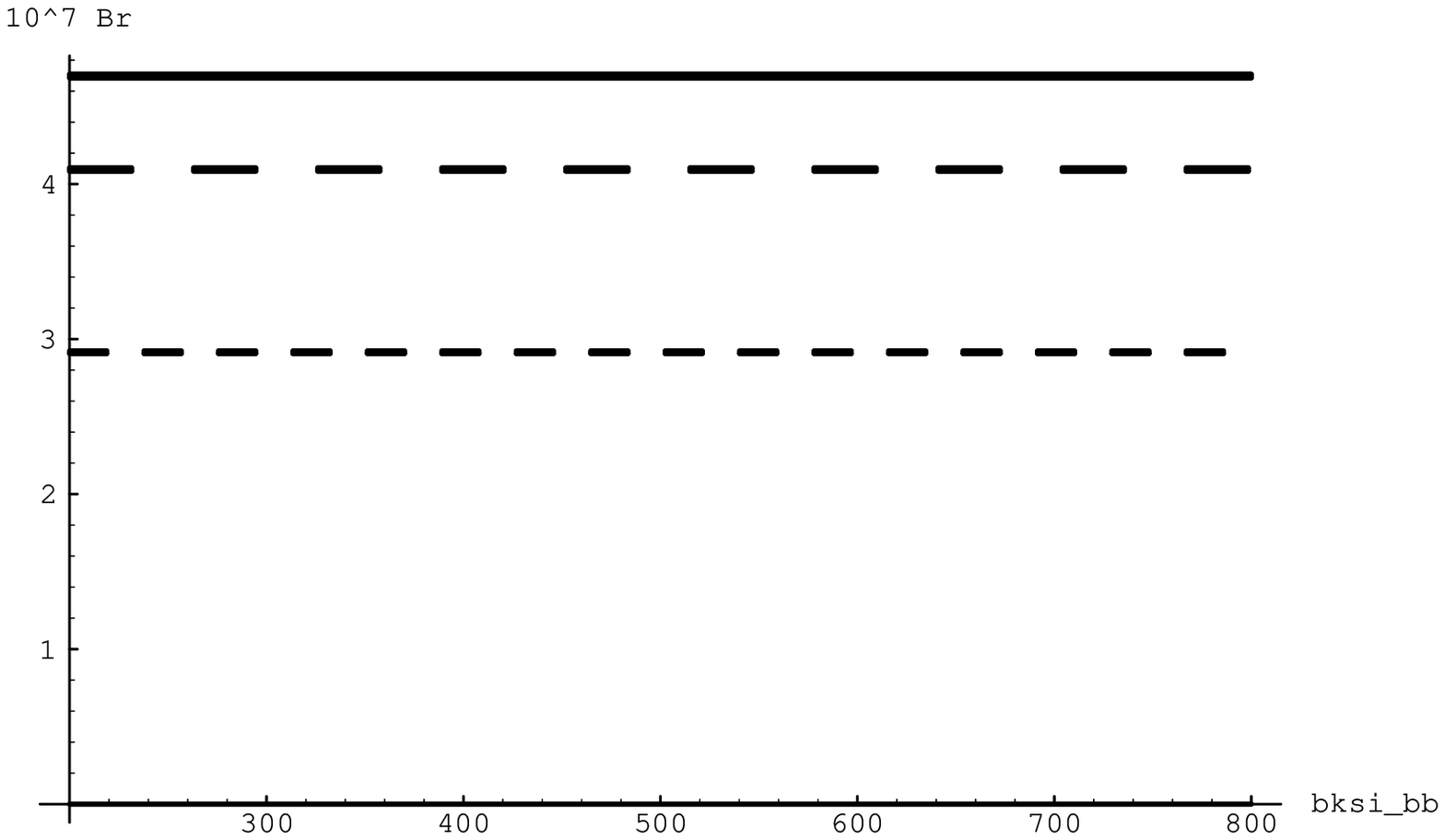}
\vskip -1.5truein
\caption[]{$Br$ as a function of the coupling $\bar{\xi}_{N,bb}^{D}$ 
for fixed $m_{H^{\pm}}=500\, GeV$ at the region $|r_{tb}|<<1$.
Here solid lines correspond to the scale $\mu=2.5\,\, GeV$, 
dashed lines to $\mu=5\,\, GeV$ and small dashed lines to $\mu=m_W\,\, GeV$. 
The lines corresponding to the SM coincides with the lines we present here.}
\label{brmh500kbb3a}
\end{figure}

\begin{figure}[htb]
\centering
\vskip -1.5truein
\epsfxsize=3.8in
\leavevmode\epsffile{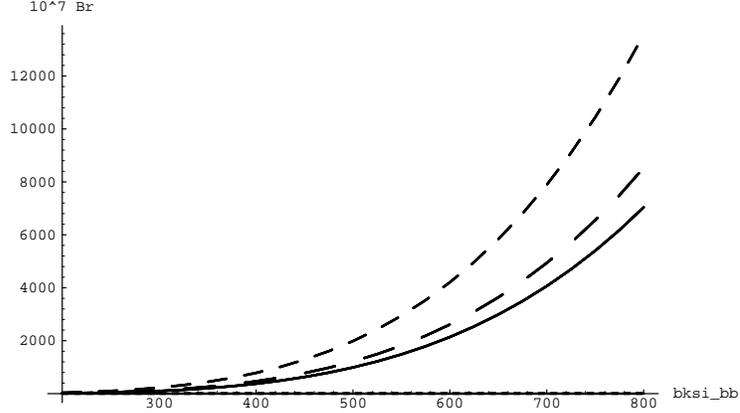}
\vskip -1.5truein
\caption[]{The same as Fig 5 but at the region $r_{tb} >> 1$.}
\label{brmh500kbb3b}
\end{figure}

After these preliminary remarks, let us start with our numerical analysis.
In this section we study the dependencies of the $Br$ and the ratio 
$R=|A^{+}|^2/|A^{-}|^2$ on the selected parameters of the model III
($\bar{\xi}^{U}_{N tt}$,  $\bar{\xi}^{D}_{N bb}$ and $m_{H^{\pm}}$) and 
the QCD scale $\mu$. 
In  figs.~\ref{brbb603mha} and \ref{brbb603mhb} 
(~\ref{brbb903mha} and ~\ref{brbb903mhb}) we plot the $Br$ 
of the decay $B_s\rightarrow \gamma\gamma$ with respect to the charged 
Higgs mass $m_{H^{\pm}}$ for the fixed value of 
$\bar{\xi}_{N,bb}^{D}=60\, m_b$ ($\bar{\xi}_{N,bb}^{D}=90\, m_b$) at 
three different $\mu$ scales ($\mu=2.5\,, 5\,,m_W \, )GeV$.
Fig.~\ref{brbb603mha} (\ref{brbb903mha}) represents the case where
the ratio $|r_{tb}|=|\frac{\bar{\xi}_{N,tt}^{U}}{\bar{\xi}_{N,bb}^{D}}| << 1$
and shows that the $Br$ obtained in the model III almost coincides with the 
one calculated in the SM. The suppression of the contribution coming from 
the charged Higgs boson can be seen from eqs.~(\ref{CoeffH}) and
\ref{Amplitudes1} with the choice $|r_{tb}|<< 1$. 
However in fig.~\ref{brbb603mhb} (\ref{brbb903mhb}), the $Br$ is presented 
for the case $r_{tb}>> 1$. It is seen that there is an extreme
enhancement of the $Br$, especially for the small values of $m_{H^{\pm}}$. 
Note that this is similar to the result coming from the choice 
of $tan\beta < 1$ in the model II \cite{gudalil}. 
In  figs.~\ref{brmh500kbb3a} ~(\ref{brmh500kbb3b}) 
we present $\bar{\xi}_{N,bb}^{D}$ dependence of 
the $Br$ at the fixed value of $m_{H^{\pm}}=500\, GeV$  for $|r_{tb}| << 1$
($r_{tb} >> 1$). In the region $|r_{tb}|<< 1$, the $Br$ is nonsensitive to
the coupling $\bar{\xi}_{N,bb}^{D}$ and almost coincides with the SM value,
however for $r_{tb} >> 1$, it increases with the increasing 
$\bar{\xi}_{N,bb}^{D}$. 
Note that the $Br$ is sensitive to the scale $\mu$. For $|r_{tb}|<< 1$, it
increases with decreasing $\mu$ 
(figs.~\ref{brbb603mha}, \ref{brbb903mha} and \ref{brmh500kbb3a} ) 
similar to the model II \cite{gudalil}. For $r_{tb} >> 1$, decreasing the 
scale $\mu$ causes the $Br$ to decrease.  
(figs.~\ref{brbb603mhb}, \ref{brbb903mhb} and \ref{brmh500kbb3b}).

At this stage we would like to estimate the upper limit of 
$\bar{\xi}_{N,bb}^{D}$ for $r_{tb} >> 1$ by using the present experimental 
result $Br(B_{s}\rightarrow \gamma\gamma)\leq 1.48 \cdot 10^{-4}$.
By choosing the lower limit for the mass $m_{H^{\pm}}\sim 480\, GeV$ \cite{gudalil}
at the scale $\mu=2.5 \, GeV$, we get $\bar{\xi}_{N,bb}^{D}< 90\, m_b$.
It is interesting to note that the increasing value of $m_{H^{\pm}}$ makes 
the restriction region for $\bar{\xi}_{N,bb}^{D}$ smaller.

Since the two photon system can be in a CP-even and CP-odd state,
$B_s\rightarrow \gamma \gamma$ decay allows us to study CP violating
effects. 
In the rest frame of the $B_s$ meson, the $CP=-1$ amplitude $A^{-}$
is proportional to the perpendicular spin polarization
$\vec{\epsilon_1}\times\vec{\epsilon_2}$, and the $CP=1$ amplitude $A^{+}$ is
proportional to the parallel spin polarization
$\vec{\epsilon_1}.\vec{\epsilon_2}$. 
The ratio $R$ is informative to search for CP violating 
effects in $B_{s}\rightarrow\gamma\gamma$ decays and it has been studied 
before in the literature in the framework of the 2HDM
without \cite{aliev} and with \cite{gudalil} QCD corrections. 

\begin{figure}[htb]
\centering
\vskip -1.5truein
\epsfxsize=3.8in
\leavevmode\epsffile{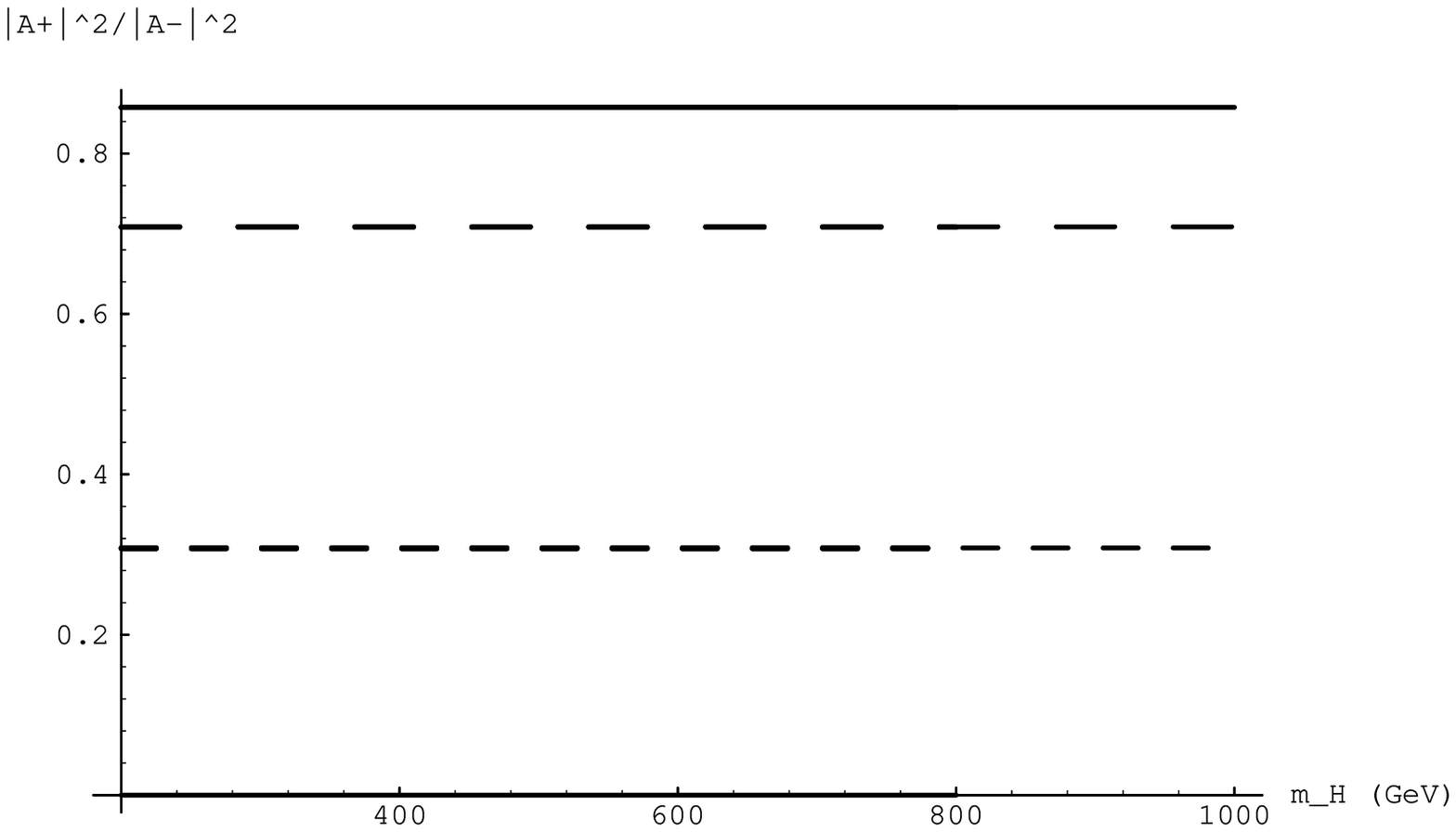}
\vskip -1.5truein
\caption[]{$R$ ratio as a function of the mass $m_{H^{\pm}}$ 
for fixed $\bar{\xi}_{N,bb}^{D}=60 \, m_b$ at the region $|r_{tb}|<<1$.
Here solid lines correspond to the scale $\mu=2.5\,\, GeV$, 
dashed lines to $\mu=5\,\, GeV$ and small dashed lines to $\mu=m_W\,\, GeV$. 
The lines corresponding to the SM coincides with the lines we present here.} 
\label{cpbb603mha}
\end{figure}

\begin{figure}[htb]
\vskip -1.5truein
\centering
\epsfxsize=3.8in
\leavevmode\epsffile{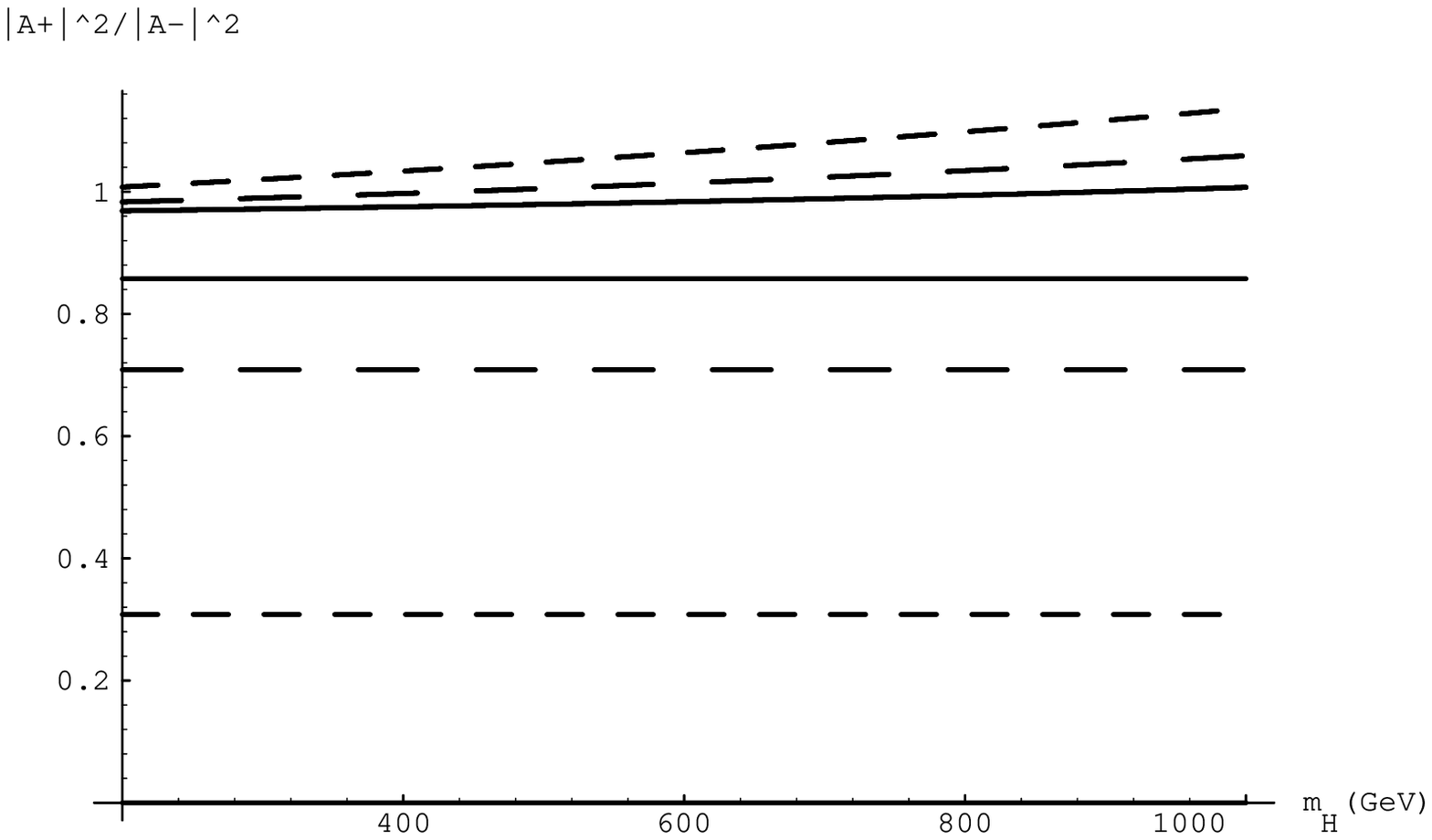}
\vskip -1.5truein
\caption[]{The same as Fig~7, but at the region $r_{tb} >> 1$.
Here solid curves (lines) correspond to the scale $\mu=2.5\,\, GeV$, 
dashed curves (lines) to $\mu=5\,\, GeV$ and small dashed curves (lines)
 to $\mu=m_W\,\, GeV$ for model III (SM).}
\label{cpbb603mhb}
\end{figure}

\begin{figure}[htb]
\vskip -1.5truein
\centering
\epsfxsize=3.8in
\leavevmode\epsffile{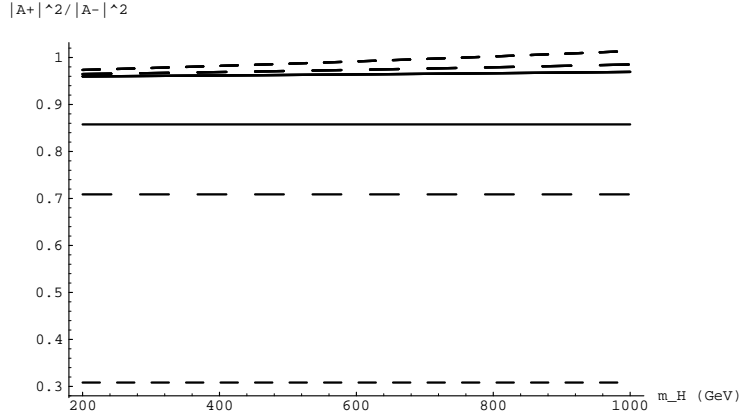}
\vskip -1.5truein
\caption[]{The same as Fig~8 but for fixed 
$\bar{\xi}_{N,bb}^{D}=100 \, m_b$.}
\label{cpbb1003mhb}
\end{figure}

\begin{figure}[htb]
\vskip -1.5truein
\centering
\epsfxsize=3.8in
\leavevmode\epsffile{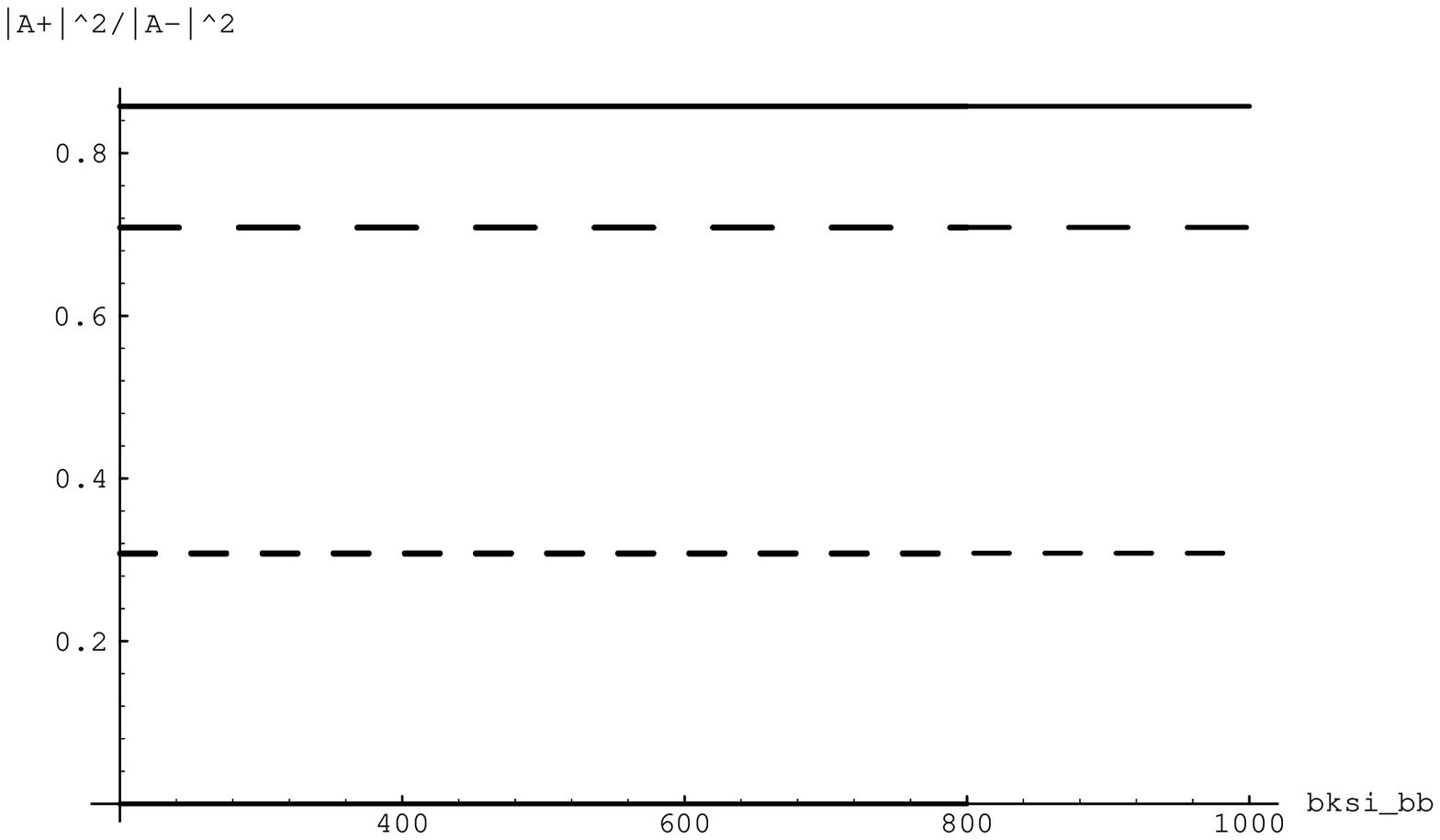}
\vskip -1.5truein
\caption[]{$R$ ratio as a function of the coupling $\bar{\xi}_{N,bb}^{D}$ 
for fixed $m_{H^{\pm}}=500\, GeV$ at the region $|r_{tb}|<<1$.
Here solid lines correspond to the scale $\mu=2.5\,\, GeV$, 
dashed lines to $\mu=5\,\, GeV$ and small dashed lines to $\mu=m_W\,\, GeV$. 
The lines corresponding to the SM coincides with the lines we present here.} 
\label{cpmh500kbb3a}
\end{figure}

\begin{figure}[htb]
\vskip -1.5truein
\centering
\epsfxsize=3.8in
\leavevmode\epsffile{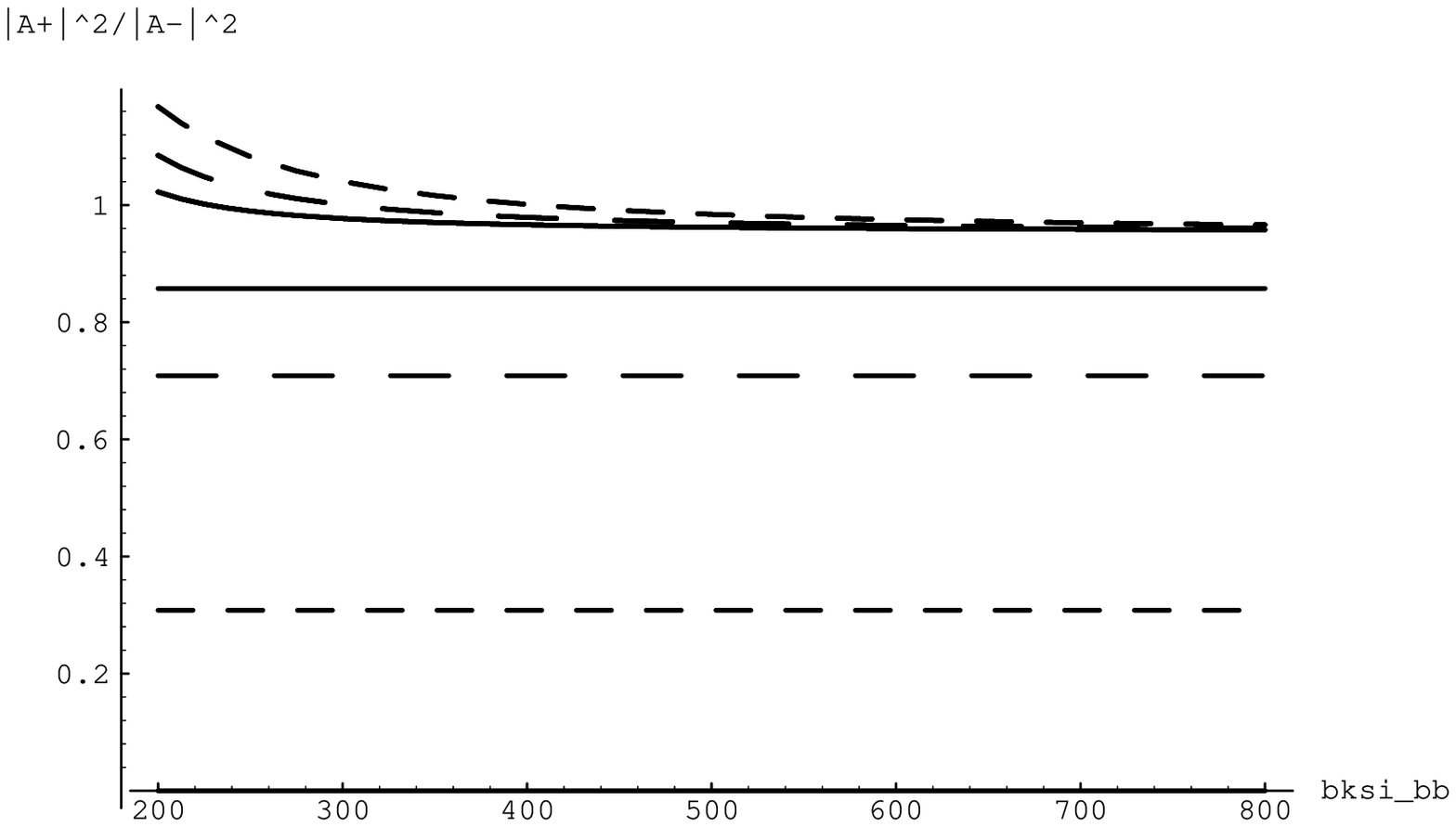}
\vskip -1.5truein
\caption[]{The same as Fig 7, but at the region $r_{tb} >> 1$.
Here solid curves (lines) correspond to the scale $\mu=2.5\,\, GeV$, 
dashed lines to $\mu=5\,\, GeV$ and small dashed lines to $\mu=m_W\,\, GeV$
for model III (SM).} 
\label{cpmh500kbb3b}
\end{figure}

Now, we present the dependence of the ratio $R=|A^{+}|^2/|A^{-}|^2$ 
on the selected parameters of model III, displayed in a series of 
figures (~\ref{cpbb603mha} - \ref{cpmh500kbb3b}).
In figs.~\ref{cpbb603mha} and ~\ref{cpbb603mhb} 
we plot the dependence of $R$ on $m_{H^{\pm}}$ for fixed 
$\bar{\xi}_{N,bb}^{D}= 60\, m_b$ and three different $\mu$ scales, 
$(2.5\,,5\,, m_{W})\,\, GeV$. The $R$ ratio is almost nonsensitive to 
$m_{H^{\pm}}$ for $|r_{tb}|<<1$ (fig.~\ref{cpbb603mha}). However, it is 
enhanced with the increasing value of $m_{H^{\pm}}$ 
($m_{H^{\pm}}\le 1000\, GeV$) 
for $r_{tb}>>1$ (fig.~\ref{cpbb603mhb}). Decreasing the scale $\mu$ weakens 
the dependence of the ratio $R$ on $m_{{H}^{\pm}}$ and the contribution of 
the charged Higgs bosons to the $R$ ratio becomes small similar to the 
model II \cite{gudalil}. Further, it becomes less dependent to $m_{H^{\pm}}$ 
with increasing 
$\bar{\xi}_{N,bb}^{D}$ (fig ~\ref{cpbb603mhb}, ~\ref{cpbb1003mhb}).

Fig.~\ref{cpmh500kbb3a} and \ref{cpmh500kbb3b} show the dependence 
of $R$ on $\bar{\xi}_{N,bb}^{D}$ for fixed $m_{{H}^{\pm}}=500\,\,GeV$. 
Like the previous case, $R$ dependence to the coupling
$\bar{\xi}_{N,bb}^{D}$ is extremely weak for $|r_{tb}| << 1$.
(fig.~\ref{cpmh500kbb3a}). For $r_{tb} >> 1$ the ratio $R$ increases
with decreasing $\bar{\xi}_{N,bb}^{D}$ (fig ~\ref{cpmh500kbb3b}).  
Note that, this ratio can exceed one unlike the SM case. The same 
situation appears in model II also \cite{gudalil}.

The ratio $R$ is quite sensitive to QCD corrections and it is 
enhanced with decreasing scale $\mu$ for the SM and $|r_{tb}| << 1$ 
(figs.~\ref{cpbb603mha}, \ref{cpmh500kbb3a}). However, this ratio decreases
for $r_{tb} >> 1$ (figs.~\ref{cpbb603mhb},~\ref{cpbb1003mhb}, 
~\ref{cpmh500kbb3b}). We observe, that the smaller the value of $m_{H^{\pm}}$ 
(the larger the value of $\bar{\xi}_{N,bb}^{D}$), the less dependent is 
the ratio on $\mu$. This strong  $\mu$ dependence makes the  
analysis of the model III parameters $m_{H^{\pm}}$ and $\bar{\xi}^{U,D}$ for 
the given experimental value of the ratio $R$ to be difficult, especially 
for the case $|r_{tb}| << 1$. However, we believe, that the strong $\mu$ 
dependence will be reduced with the addition the of next to leading order 
(NLO) calculation. This requires a computation of finite parts of many two
loop diagrams, and divergent part of three-loop diagrams. This lies beyond
the scope of the present work and it has to be done to reduce the 
uncertainity coming from the scale $\mu$. Fortunately, the choice of 
$\mu=m_b/2$ in the LLog approximation reproduces effectively the NLO result 
for the $b\rightarrow s\gamma$ decay and one can suggest that it may also 
work for the $b\rightarrow s\gamma\gamma$ decay. In any case, the analysis 
on the model III parameters will be more reliable after NLO calculations 
are done.

We complete this section by
taking the $O_{7}$ type long distance effects ($LD_{O_{7}}$)
for both the $Br$ and the ratio $R$ into account.
The $LD_{O_{7}}$ contribution to the CP-odd $A^{-}$ and CP-even $A^{+}$ 
amplitudes has been calculated with the help of the Vector Meson
Dominance model (VMD) \cite{Gud} and it was shown, that the influence on 
the amplitudes was destructive. With the addition of the $LD_{O_{7}}$ 
effects, the amplitudes entering the $Br$ and $R$ ratio are now given as
\begin{eqnarray}
A^{+}&=& A^{+}_{SD}+A^{+}_{LD_{O_{7}}}\nonumber \,\, , \\
A^{-}&=& A^{-}_{SD}+A^{-}_{LD_{O_{7}}} \,\, , 
\label{Amptot}
\end{eqnarray}
where $A^{\pm}_{SD}$ are the short distance amplitudes we took into account
in the previous sections (eq.(\ref{finamp})). The $LD_{O_{7}}$ amplitudes 
$A^{\pm}_{LD_{O_{7}}}$ are defined as \cite{Gud} 
\begin{eqnarray}
A^{+}_{LD_{O_7}}&=&
-\sqrt{2} \frac{\alpha_{em} G_F}{\pi} \bar{F_1}(0) f_{\phi}(0) 
\lambda_t \frac{m_b (m_{B_s}^2-m_{\phi}^2)}{3 m_{\phi} m_{B_{s}}^{2}} 
C_{7}^{eff}(\mu)  \, \, , \nonumber\\
A^{-}_{LD_{O_7}}&=&
\sqrt{2} \frac{\alpha_{em} G_F}{\pi} \bar{F_1}(0) f_{\phi}(0) 
\lambda_t \frac{m_b}{3 m_{\phi}} C_{7}^{eff}(\mu) \, \, ,
\label{AmpLD}
\end{eqnarray}
where $f_{\phi}(0)=0.18\,\, GeV$ is the decay constant of $\phi$ meson
at zero momentum, 
$\bar{F}_{1}(0)$ is the extrapolated $B_{s}\rightarrow \phi$ form factor
(for details see \cite{Gud}).
Note that we neglect the contribution of operator $O'_{7}$ 
in eq.(~\ref{AmpLD}) since the coefficient $C_{7}^{\prime eff}(\mu)$ 
is negligible compared to the coefficient $C_{7}^{eff}(\mu)$ (see
the discussion given in appendix).

In figs. (~\ref{brbb603mhbLD} -~\ref{cpmh500kbb3bLD} )
we present the $m_{H^{\pm}}$ and $\bar{\xi}_{N,bb}^{D}$ dependencies of
the $Br$ and the ratio $R$  with the addition of $LD_{O_{7}}$ effects. 
Here we use $\bar{F}_{1}(0)=0.16$ \cite{Gud}. 
The $Br$ decreases with the addition of $LD_{O_{7}}$ effects,
since the effect is destructive.  
The $\mu$ scale uncertanity of the $Br$ is smaller compared to the case 
where LD effect is not included.

It can be shown, that the value of the ratio $R$ also decreases
with the addition of $LD_{O_{7}}$ effects for the SM case. 
(figs. ~\ref{cpbb603mhaLD} - \ref{cpmh500kbb3bLD}). However, while 
$m_{H^{\pm}}$ is increasing or $\bar{\xi}_{N,bb}^{D}$ is decreasing, 
the effect of the $LD_{O_{7}}$ contribution on the ratio $R$ is increasing 
for $r_{tb}>>1$ (figs.~\ref{cpbb603mhbLD}, ~\ref{cpmh500kbb3bLD}).

There are still non-perturbative effects which can come from the formation 
of $c\bar{c}$ bound states. However these states are far off-shell and do 
not give significant contribution to the decay rate \cite{Soni}. 
For example the chain process 
$B_s\rightarrow \phi\psi\rightarrow\phi\gamma\rightarrow\gamma\gamma$ is
estimated and it is found to be at most $1\%$ of the branching ratio
$Br(B_s\rightarrow\gamma\gamma)_{SD+LD_{O_7}}$ \cite{gudil} 

Besides the strong $\mu$ dependence there is another uncertainity coming 
from the choice of bound state parameters $m_b^{eff}$ and $\bar{\Lambda}_s$.
It follows that the larger $m_b^{eff}$ (smaller $\bar{\Lambda}_s$),
the larger $Br$ and $R$ ratio. Here the enhacement of the $Br$ is caused 
by the $\frac{1}{\bar{\Lambda}_s}$ dependence in amplitudes. 

In conclusion, we analyse the selected model III parameters 
( $\bar{\xi}_{N,bb}^{D}$, $\bar{\xi}_{N,tt}^{U}$, $m_{H^{\pm}}$) and QCD 
scale $\mu$ dependencies of the $Br$ and $R$ ratio for the decay
$B_s\rightarrow\gamma\gamma$. We predicted the upper bound 
for $\bar{\xi}_{N,bb}^{D}$, $\bar{\xi}_{N,bb}^{D} \le 90\, m_b$ in the
case of $r_{tb} >> 1$, using the constraints for  $\bar{\xi}_{N,bb}^{D}$, 
$\bar{\xi}_{N,tt}^{U}$ \cite{alil} and the experimental upper limit for 
the $Br$ of the decay underconsideration. We obtain that the strong 
enhancement of the $Br$ is possible in the framework of the model III.

\begin{appendix}
\section{Appendix \\ The operator basis and the Wilson coefficients 
for the decay $b\rightarrow s \gamma\gamma$ in the model III}

The operator basis is the same as the one used for the $b\rightarrow s
\gamma$ decay in the  model III \cite{alil} and 
$SU(2)_L\times SU(2)_R\times U(1)$ extensions of the SM \cite{cho}:
\begin{eqnarray}
 O_1 &=& (\bar{s}_{L \alpha} \gamma_\mu c_{L \beta})
               (\bar{c}_{L \beta} \gamma^\mu b_{L \alpha}), \nonumber   \\
 O_2 &=& (\bar{s}_{L \alpha} \gamma_\mu c_{L \alpha})
               (\bar{c}_{L \beta} \gamma^\mu b_{L \beta}),  \nonumber   \\
 O_3 &=& (\bar{s}_{L \alpha} \gamma_\mu b_{L \alpha})
               \sum_{q=u,d,s,c,b}
               (\bar{q}_{L \beta} \gamma^\mu q_{L \beta}),  \nonumber   \\
 O_4 &=& (\bar{s}_{L \alpha} \gamma_\mu b_{L \beta})
                \sum_{q=u,d,s,c,b}
               (\bar{q}_{L \beta} \gamma^\mu q_{L \alpha}),   \nonumber  \\
 O_5 &=& (\bar{s}_{L \alpha} \gamma_\mu b_{L \alpha})
               \sum_{q=u,d,s,c,b}
               (\bar{q}_{R \beta} \gamma^\mu q_{R \beta}),   \nonumber  \\
 O_6 &=& (\bar{s}_{L \alpha} \gamma_\mu b_{L \beta})
                \sum_{q=u,d,s,c,b}
               (\bar{q}_{R \beta} \gamma^\mu q_{R \alpha}),  \nonumber   \\  
 O_7 &=& \frac{e}{16 \pi^2}
          \bar{s}_{\alpha} \sigma_{\mu \nu} (m_b R + m_s L) b_{\alpha}
                {\cal{F}}^{\mu \nu},                             \nonumber       \\
 O_8 &=& \frac{g}{16 \pi^2}
    \bar{s}_{\alpha} T_{\alpha \beta}^a \sigma_{\mu \nu} (m_b R + m_s L)  
          b_{\beta} {\cal{G}}^{a \mu \nu} \nonumber \,\, , \\  
 O_9 &=& (\bar{s}_{L \alpha} \gamma_\mu c_{L \beta})
               (\bar{c}_{R \beta} \gamma^\mu b_{R \alpha}), \nonumber   \\
 O_{10} &=& (\bar{s}_{L \alpha} \gamma_\mu c_{L \alpha})
(\bar{c}_{R \beta} \gamma^\mu b_{R \beta}),
\label{op1}
\end{eqnarray}
and the second operator set $O'_{1} - O'_{10}$ which are 
flipped chirality partners of $O_{1} - O_{10}$:
\begin{eqnarray}
 O'_1 &=& (\bar{s}_{R \alpha} \gamma_\mu c_{R \beta})
               (\bar{c}_{R \beta} \gamma^\mu b_{R \alpha}), \nonumber   \\
 O'_2 &=& (\bar{s}_{R \alpha} \gamma_\mu c_{R \alpha})
               (\bar{c}_{R \beta} \gamma^\mu b_{R \beta}),  \nonumber   \\
 O'_3 &=& (\bar{s}_{R \alpha} \gamma_\mu b_{R \alpha})
               \sum_{q=u,d,s,c,b}
               (\bar{q}_{R \beta} \gamma^\mu q_{R \beta}),  \nonumber   \\
 O'_4 &=& (\bar{s}_{R \alpha} \gamma_\mu b_{R \beta})
                \sum_{q=u,d,s,c,b}
               (\bar{q}_{R \beta} \gamma^\mu q_{R \alpha}),   \nonumber  \\
 O'_5 &=& (\bar{s}_{R \alpha} \gamma_\mu b_{R \alpha})
               \sum_{q=u,d,s,c,b}
               (\bar{q}_{L \beta} \gamma^\mu q_{L \beta}),   \nonumber  \\
 O'_6 &=& (\bar{s}_{R \alpha} \gamma_\mu b_{R \beta})
                \sum_{q=u,d,s,c,b}
               (\bar{q}_{L \beta} \gamma^\mu q_{L \alpha}),  \nonumber   \\  
 O'_7 &=& \frac{e}{16 \pi^2}
          \bar{s}_{\alpha} \sigma_{\mu \nu} (m_b L + m_s R) b_{\alpha}
                {\cal{F}}^{\mu \nu},                             \nonumber       \\
 O'_8 &=& \frac{g}{16 \pi^2}
    \bar{s}_{\alpha} T_{\alpha \beta}^a \sigma_{\mu \nu} (m_b L + m_s R)  
          b_{\beta} {\cal{G}}^{a \mu \nu}, \nonumber \\ 
 O'_9 &=& (\bar{s}_{R \alpha} \gamma_\mu c_{R \beta})
               (\bar{c}_{L \beta} \gamma^\mu b_{L \alpha})\,\, , \nonumber   \\
 O'_{10} &=& (\bar{s}_{R \alpha} \gamma_\mu c_{R \alpha})
(\bar{c}_{L \beta} \gamma^\mu b_{L \beta})\,\, ,
\label{op2}
\end{eqnarray}
where  
$\alpha$ and $\beta$ are $SU(3)$ colour indices and
${\cal{F}}^{\mu \nu}$ and ${\cal{G}}^{\mu \nu}$
are the field strength tensors of the electromagnetic and strong
interactions, respectively.

In the calculations, we take only the charged Higgs contributions into 
account and neglect the effects of neutral Higgs bosons. At this stage
we would like to give the reasons by using the restrictions to the 
effective Wilson coefficient $C_7^{eff}$ coming from the CLEO data in 
the process $B\rightarrow K^* \gamma$. (see section 2 and \cite{alil} also)

The neutral bosons $H_0$, $H_1$ and $H_2$ are defined  in terms of the
mass eigenstates $\bar{H}_0$ ,$h_0$ and $A_0$ as 
\begin{eqnarray}
H_{0}&=&( \bar{H}_{0}cos \alpha - h_0 sin\alpha)+v \nonumber \, ,\\ 
H_{1}&=&( h_{0}cos \alpha + \bar{H}_0 sin\alpha) \nonumber \, ,\\ 
H_{2}&=&A_0 \,\,,
\label{neutrbos}
\end{eqnarray}
where $\alpha$ is the mixing angle and $v$ is proportional to the vacuum 
expectation value of the doublet $\phi_1$ (eq. (\ref{choice2})). Here we
assume that the massess of neutral Higgs bosons $h_0$ and $A_0$ are heavy
compared to the b-quark mass. The neutral Higgs scalar $h_0$ and pseduscalar 
$A_0$ give contribution only to $C_7$ for our process. With the choice of 
$\alpha=0$, $C_7^{h_0}$ and $C_7^{A_0}$ can be calculated at $m_W$ level as
\begin{eqnarray}
C_7^{h_0}(m_W)&=& (V_{tb} V^{*}_{ts} )^{-1}\sum_{i=d,s,b} \bar{\xi}^{D}_{N,bi} 
\,\,\bar{\xi}^{D}_{N,is}\,  \frac{Q_i}{8\, m_i\, m_b}
\nonumber \,\,, \\ 
C_7^{A_0}(m_W)&=& (V_{tb} V^{*}_{ts} )^{-1}\sum_{i=d,s,b} \bar{\xi}^{D}_{N,bi} 
\,\, \bar{\xi}^{D}_{N,is}\, \frac{Q_i}{8\, m_i\, m_b}
\,\, ,
\label{c7A0h0}
\end{eqnarray}
where $m_i$ and $Q_i$ are the masses and charges of the down quarks 
($i=d,\,s,\,b$) respectively. Here we used the redefinition
\begin{eqnarray}
\xi^{U,D}=\sqrt{\frac{4 G_{F}}{\sqrt{2}} \,\, \bar{\xi}^{U,D}}\,\, .
\label{ksidefn}
\end{eqnarray} 
Eq. (\ref{c7A0h0}) shows that neutral Higgs bosons can give a large 
contribution to $C_7$, which does not respect the CLEO data \cite{cleo}. 
Here, we make an assumption that the couplings 
$\bar{\xi}^{D}_{N,is}$($i=d,s,b)$ and $\bar{\xi}^{D}_{N,db}$ are 
negligible to be able to reach the conditions 
$\bar{\xi}^{D}_{N,bb} \,\bar{\xi}^{D}_{N,is} <<1$ and 
$\bar{\xi}^{D}_{N,db} \,\bar{\xi}^{D}_{N,ds} <<1$.
These choices permit us to neglect the neutral Higgs effects.

Denoting the Wilson coefficients for the SM with $C_{i}^{SM}(m_{W})$ and the
additional charged Higgs contribution with $C_{i}^{H}(m_{W})$, 
we have the initial values for the first set of operators 
(eq.(~\ref{op1})) (\cite{alil} and references within) 
\begin{eqnarray}
C^{SM}_{1,3,\dots 6,9,10}(m_W)&=&0 \nonumber \, \, , \\
C^{SM}_2(m_W)&=&1 \nonumber \, \, , \\
C_7^{SM}(m_W)&=&\frac{3 x^3-2 x^2}{4(x-1)^4} \ln x+
\frac{-8x^3-5 x^2+7 x}{24 (x-1)^3} \nonumber \, \, , \\
C_8^{SM}(m_W)&=&-\frac{3 x^2}{4(x-1)^4} \ln x+
\frac{-x^3+5 x^2+2 x}{8 (x-1)^3}\nonumber \, \, , \\ 
C^{H}_{1,\dots 6,9,10}(m_W)&=&0 \nonumber \, \, , \\
C_7^{H}(m_W)&=&\frac{1}{m_{t}^2} \,
(\bar{\xi}^{U}_{N,tt}+\bar{\xi}^{U}_{N,tc}
\frac{V_{cs}^{*}}{V_{ts}^{*}}) \, (\bar{\xi}^{U}_{N,tt}+\bar{\xi}^{U}_{N,tc}
\frac{V_{cb}}{V_{tb}}) F_{1}(y)\nonumber  \, \, , \\
&+&\frac{1}{m_t m_b} \, (\bar{\xi}^{U}_{N,tt}+\bar{\xi}^{U}_{N,tc}
\frac{V_{cs}^{*}}{V_{ts}^{*}}) \, (\bar{\xi}^{D}_{N,bb}+\bar{\xi}^{D}_{N,sb}
\frac{V_{ts}}{V_{tb}}) F_{2}(y)\nonumber  \, \, , \\
C_8^{H}(m_W)&=&\frac{1}{m_{t}^2} \,
(\bar{\xi}^{U}_{N,tt}+\bar{\xi}^{U}_{N,tc}
\frac{V_{cs}^{*}}{V_{ts}^{*}}) \, (\bar{\xi}^{U}_{N,tt}+\bar{\xi}^{U}_{N,tc}
\frac{V_{cb}}{V_{tb}})G_{1}(y)\nonumber  \, \, , \\
&+&\frac{1}{m_t m_b} \, (\bar{\xi}^{U}_{N,tt}+\bar{\xi}^{U}_{N,tc}
\frac{V_{cs}^{*}}{V_{ts}^{*}}) \, (\bar{\xi}^{D}_{N,bb}+\bar{\xi}^{U}_{N,sb}
\frac{V_{ts}}{V_{tb}}) G_{2}(y) \, \, ,
\label{CoeffH}
\end{eqnarray}
and for the second set of operators eq.~(\ref{op2}), 
\begin{eqnarray}
C^{\prime SM}_{1,\dots 10}(m_W)&=&0 \nonumber \, \, , \\
C^{\prime H}_{1,\dots 6,9,10}(m_W)&=&0 \nonumber \, \, , \\
C^{\prime H}_7(m_W)&=&\frac{1}{m_t^2} \,
(\bar{\xi}^{D}_{N,bs}\frac{V_{tb}}{V_{ts}^{*}}+\bar{\xi}^{D}_{N,ss})
\, (\bar{\xi}^{D}_{N,bb}+\bar{\xi}^{D}_{N,sb}
\frac{V_{ts}}{V_{tb}}) F_{1}(y)\nonumber  \, \, , \\
&+& \frac{1}{m_t m_b}\, (\bar{\xi}^{D}_{N,bs}\frac{V_{tb}}{V_{ts}^{*}}
+\bar{\xi}^{D}_{N,ss}) \, (\bar{\xi}^{U}_{N,tt}+\bar{\xi}^{U}_{N,tc}
\frac{V_{cb}}{V_{tb}}) F_{2}(y)\nonumber  \, \, , \\
C^{\prime H}_8 (m_W)&=&\frac{1}{m_t^2} \,
(\bar{\xi}^{D}_{N,bs}\frac{V_{tb}}{V_{ts}^{*}}+\bar{\xi}^{D}_{N,ss})
\, (\bar{\xi}^{D}_{N,bb}+\bar{\xi}^{D}_{N,sb}
\frac{V_{ts}}{V_{tb}}) G_{1}(y)\nonumber  \, \, , \\
&+&\frac{1}{m_t m_b} \, (\bar{\xi}^{D}_{N,bs}\frac{V_{tb}}{V_{ts}^{*}}
+\bar{\xi}^{D}_{N,ss}) \, (\bar{\xi}^{U}_{N,tt}+\bar{\xi}^{U}_{N,tc}
\frac{V_{cb}}{V_{tb}}) G_{2}(y) \,\, ,
\label{CoeffH2}
\end{eqnarray}
where $x=m_t^2/m_W^2$ and $y=m_t^2/m_{H^{\pm}}^2$.
The functions $F_{1}(y)$, $F_{2}(y)$, $G_{1}(y)$ and $G_{2}(y)$ are given as
\begin{eqnarray}
F_{1}(y)&=& \frac{y(7-5y-8y^2)}{72 (y-1)^3}+\frac{y^2 (3y-2)}{12(y-1)^4}
\,
ln y \nonumber  \,\, , \\ 
F_{2}(y)&=& \frac{y(5y-3)}{12 (y-1)^2}+\frac{y(-3y+2)}{6(y-1)^3}\, ln y 
\nonumber  \,\, ,\\ 
G_{1}(y)&=& \frac{y(-y^2+5y+2)}{24 (y-1)^3}+\frac{-y^2} {4(y-1)^4} \, ln y
\nonumber  \,\, ,\\ 
G_{2}(y)&=& \frac{y(y-3)}{4 (y-1)^2}+\frac{y} {2(y-1)^3} \, ln y \,\, .
\label{F1G1}
\end{eqnarray}
Note that we neglect the contributions of the internal $u$ and $c$ quarks 
compared to one due to the internal $t$ quark.

For the initial values of the Wilson coefficients in the model III  
(eqs. (\ref{CoeffH})and (\ref{CoeffH2})), we have 
\begin{eqnarray}
C^{2HDM}_{1,3,\dots 6,9,10}(m_W)&=&0 \nonumber \, \, , \\
C_2^{2HDM}(m_W)&=&1 \nonumber \, \, , \\
C_7^{2HDM}(m_W)&=&C_7^{SM}(m_W)+C_7^{H}(m_W) \nonumber \, \, , \\
C_8^{2HDM}(m_W)&=&C_8^{SM}(m_W)+C_8^{H}(m_W) \nonumber \, \, , \\ 
C^{\prime 2HDM}_{1,2,3,\dots 6,9,10}(m_W)&=&0 \nonumber \, \, , \\
C_7^{\prime 2HDM}(m_W)&=&C_7^{\prime SM}(m_W)+C_7^{\prime H}(m_W) \nonumber \, \, , \\
C_8^{\prime 2HDM}(m_W)&=&C_8^{\prime SM}(m_W)+C_8^{\prime H}(m_W) \, \, . 
\label{Coef2HDM}
\end{eqnarray}

At this stage it is possible to obtain the result for model II, in the
approximation $\frac{m_{s}}{m_{b}}\sim 0$ and 
$\frac{m_{b}^2}{m_{t}^2}\sim 0$, by making the
following replacements in the Wilson coefficients:
\begin{eqnarray}
\bar{\xi}^{U *}_{st}\bar{\xi}^{U}_{tb}&=&m_{t}^2
\frac{1}{tan^{2}\beta}\nonumber \,\, ,\\
\bar{\xi}^{U *}_{st}\bar{\xi}^{D}_{tb}&=&-m_{t} m_{b}\,\, ,
\label{replacement}
\end{eqnarray}
and taking zero for the coefficients of the flipped operator set, i.e
$C^{\prime}_{i}\rightarrow 0$. 

The evaluation of the Wilson coefficients are done by using the initial 
values $C_i^{2HDM}$ ($C_i^{\prime 2HDM}$) and  their contributions at any 
lower scale can be calculated as in the SM case \cite{alil}. 
\end{appendix}

\newpage

\begin{figure}[htb]
\vskip -1.5truein
\centering
\epsfxsize=3.8in
\leavevmode\epsffile{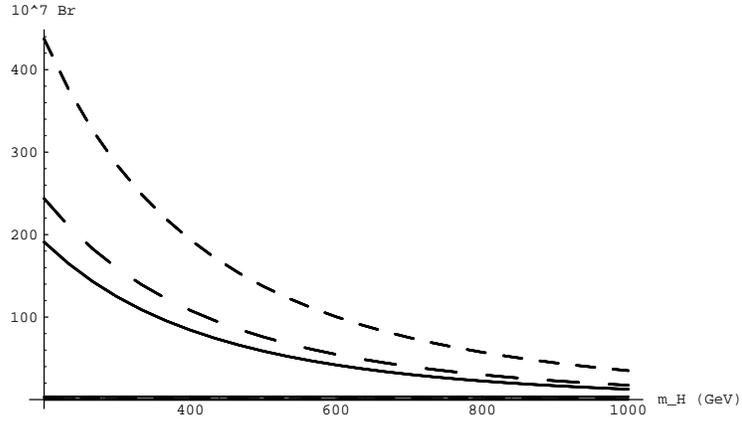}
\vskip -1.5truein
\caption[]{The same as Fig. 2 but including LD effects.
The lines corresponding to the SM coincides almost with the $m_{H^{\pm}}$ axis.}
\label{brbb603mhbLD}
\end{figure}

\begin{figure}[htb]
\vskip -1.5truein
\centering
\epsfxsize=3.8in
\leavevmode\epsffile{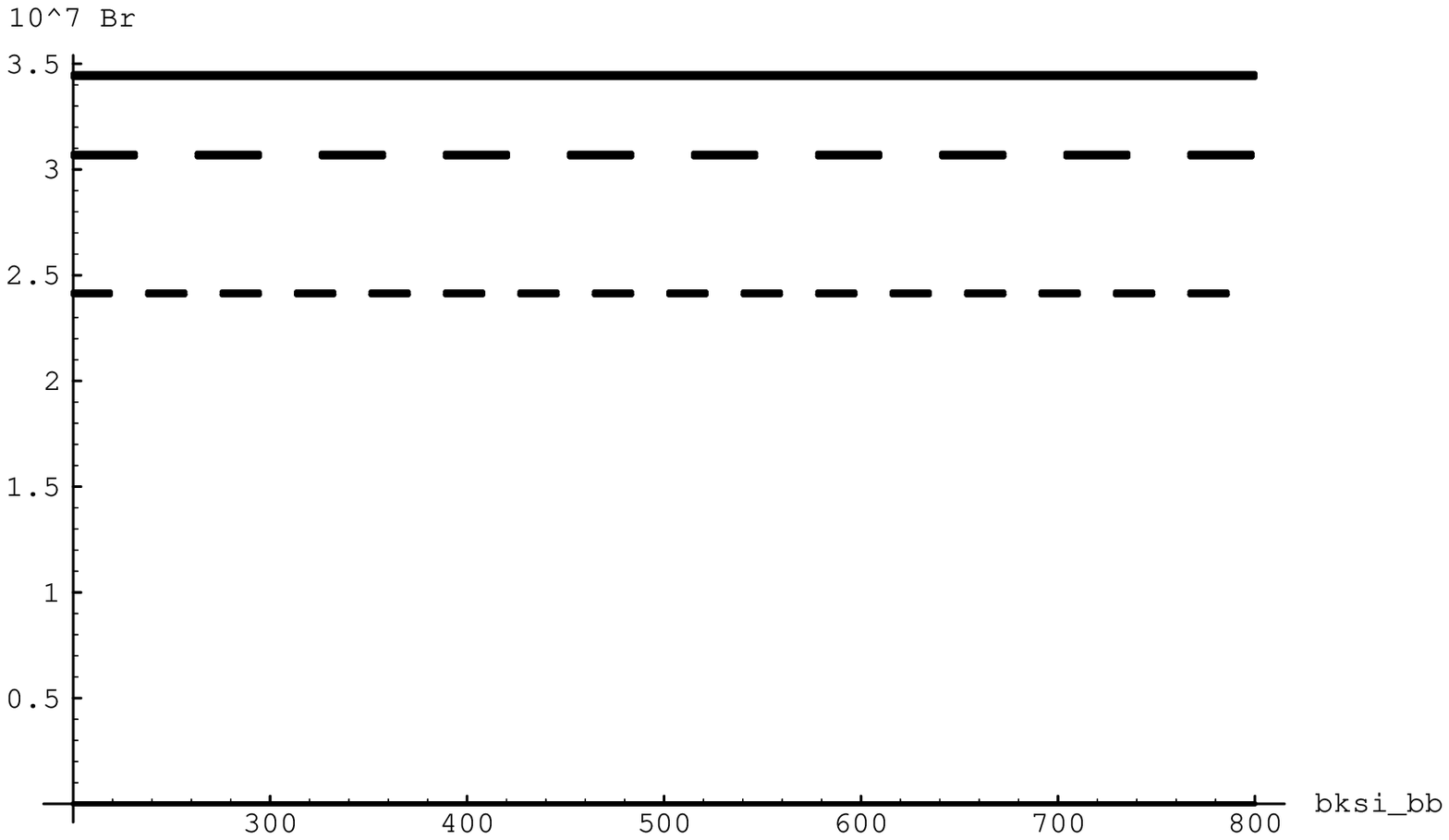}
\vskip -1.5truein
\caption[]{The same as Fig. 5 but including LD effects.}
\label{brmh500kbb3aLD}
\end{figure}

\begin{figure}[htb]
\vskip -1.5truein
\centering
\epsfxsize=3.8in
\leavevmode\epsffile{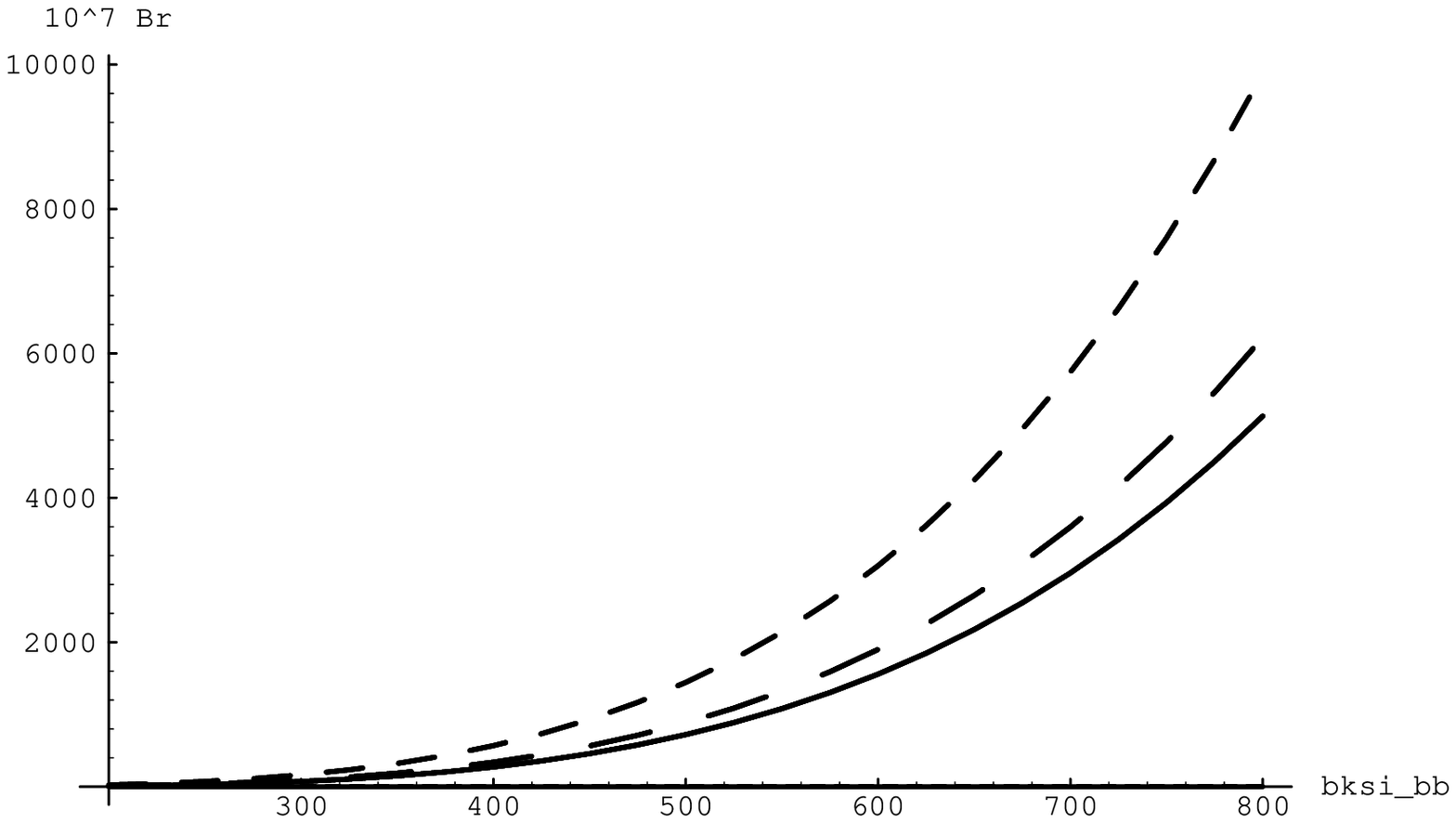}
\vskip -1.5truein
\caption[]{The same as Fig. 6 but including LD effects.}
\label{brmh500kbb3bLD}
\end{figure}

\begin{figure}[htb]
\vskip -1.5truein
\centering
\epsfxsize=3.8in
\leavevmode\epsffile{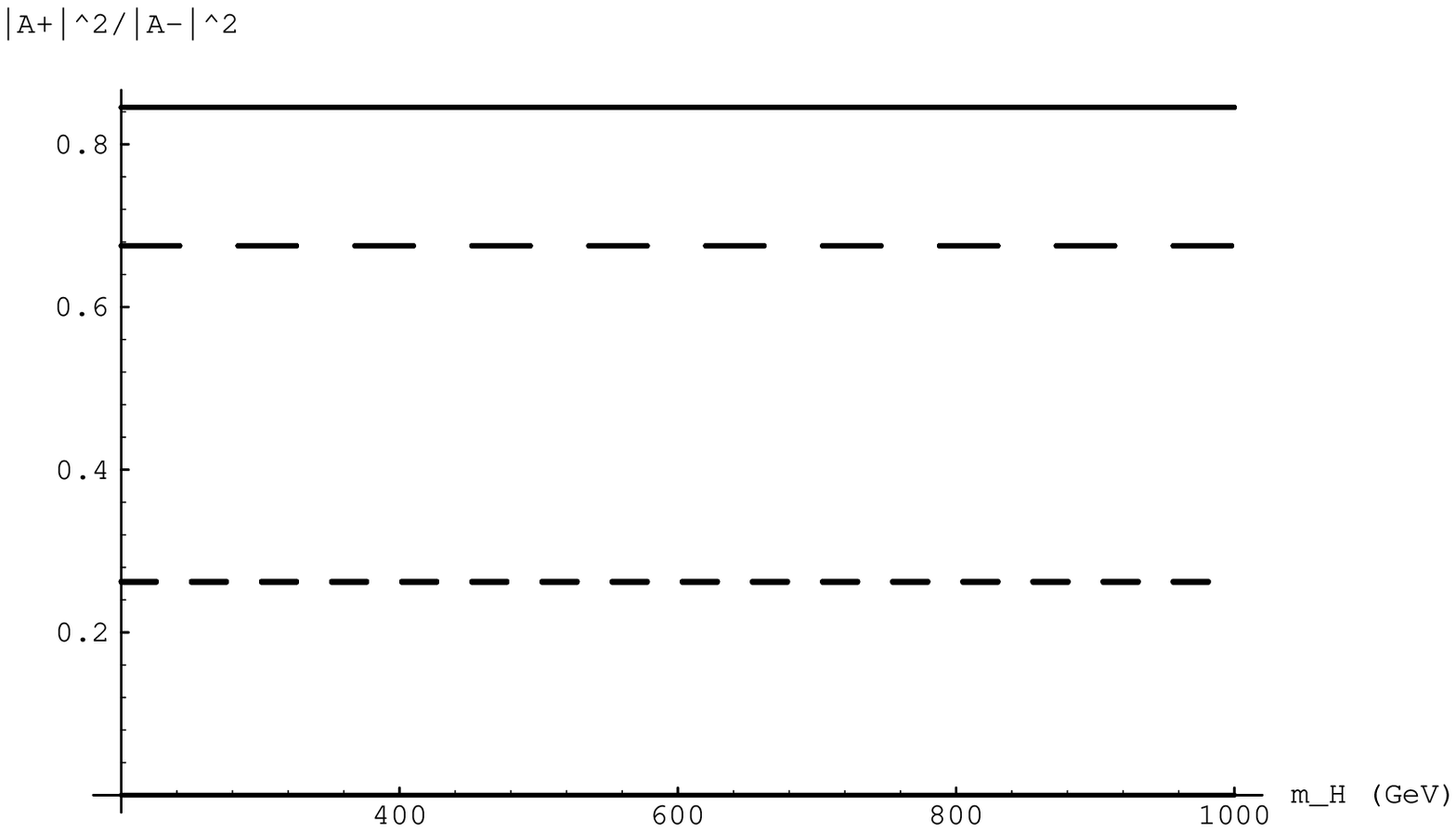}
\vskip -1.5truein
\caption[]{The same as Fig. 7 but including LD effects.}
\label{cpbb603mhaLD}
\end{figure}

\begin{figure}[htb]
\vskip -1.5truein
\centering
\epsfxsize=3.8in
\leavevmode\epsffile{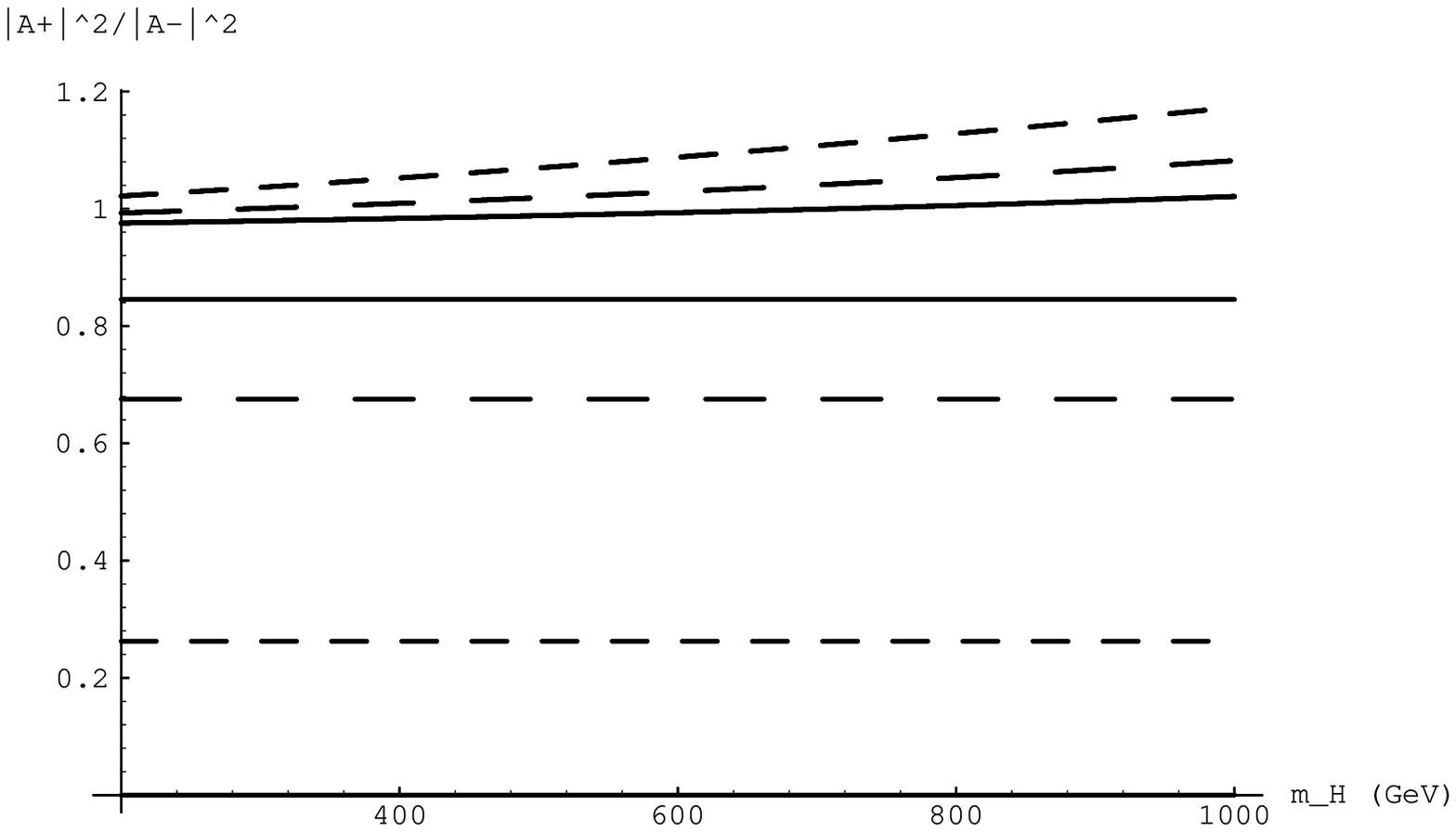}
\vskip -1.5truein
\caption[]{The same as Fig. 8 but including LD effects.}
\label{cpbb603mhbLD}
\end{figure}

\begin{figure}[htb]
\vskip -1.5truein
\centering
\epsfxsize=3.8in
\leavevmode\epsffile{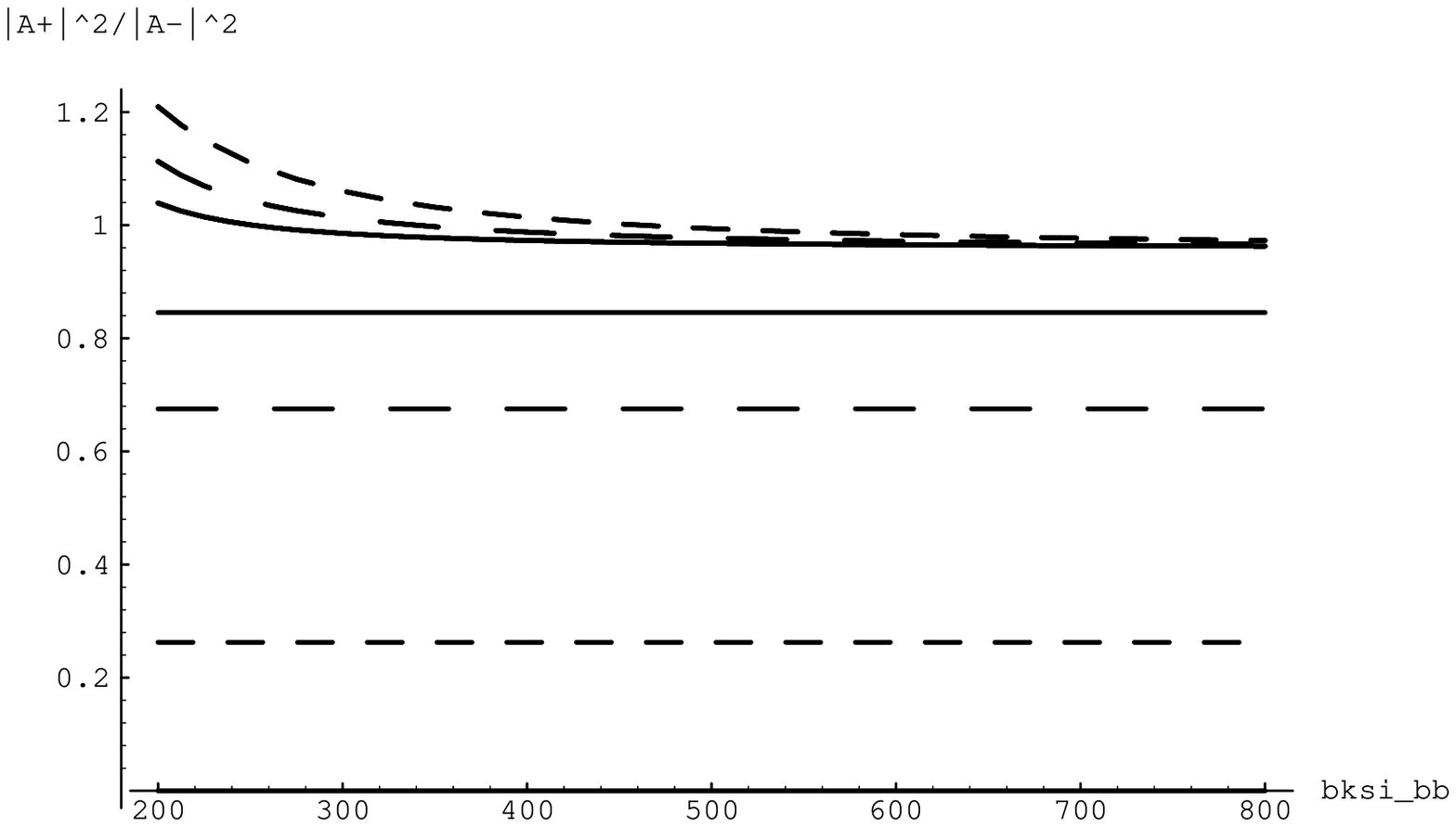}
\vskip -1.5truein
\caption[]{The same as Fig 11 , but including LD effects.}
\label{cpmh500kbb3bLD}
\end{figure}

\end{document}